\documentclass[ASNA,twocolumn]{USG} 
\usepackage{anyfontsize} %
\usepackage{colortbl}   
\usepackage{xcolor}     
\usepackage{steinmetz}
\usepackage{makecell} 
\usepackage{multirow} 
\usepackage{textcomp} 
\usepackage{pifont} 
\usepackage{xspace} 
\usepackage{appendix} 
\usepackage{framed} 
\usepackage{hyperref} 
\usepackage{tablefootnote}
\usepackage{threeparttable} 
\usepackage{color}
\usepackage{array}
\usepackage{sparklines}
\usepackage{tikz}
\usetikzlibrary{calc}
\usepackage{booktabs} 
\usepackage{tabularx} 
\usepackage{amsmath}
\usepackage{algorithm}
\usepackage{algpseudocode}
\usepackage[utf8]{inputenc}
\usepackage{listings}
\usepackage{tcolorbox}
\tcbuselibrary{listings, skins, breakable}

\definecolor{cadregreen}{RGB}{0, 128, 0}
\definecolor{vanillared}{RGB}{220, 20, 60}
\definecolor{origgray}{RGB}{128, 128, 128}
\definecolor{bgcadre}{RGB}{240, 255, 240}
\definecolor{bgvanilla}{RGB}{255, 240, 240}
\definecolor{bgorig}{RGB}{248, 248, 248}

\newcommand{\pluscadre}{%
  \makebox[0pt][l]{\textcolor{cadregreen}{\textbf{+}}\hspace{0.4em}}%
}
\newcommand{\plusvanilla}{%
  \makebox[0pt][l]{\textcolor{vanillared}{\textbf{+}}\hspace{0.4em}}%
}

\newtcblisting{methodlisting}[4][]{
  enhanced,
  breakable,
  listing only,
  colback=#3,
  colframe=#4,
  boxrule=0.8pt,
  arc=2pt,
  left=6pt,
  right=6pt,
  top=0pt,        
  bottom=0pt,     
  title={\textbf{#2}},
  fonttitle=\sffamily\normalsize,
  coltitle=#4!80!black,
  attach boxed title to top left={yshift=-2.5mm, xshift=4mm},
  boxed title style={
    colback=white,
    boxrule=0pt,
    frame hidden,
  },
  listing options={
    language=bash,
    basicstyle=\normalsize\ttfamily,
    breaklines=true,
    showstringspaces=false,
    tabsize=2,
    columns=fullflexible,
    keepspaces=true,
    escapeinside={(*}{*)},
    #1
  },
}

\newtcblisting{dockerlisting}[1]{
  listing only,
  title=#1,
  fonttitle=\bfseries\normalsize,
  colbacktitle=gray!60!white,
  coltitle=white,
  colback=gray!8,
  colframe=gray!40,
  boxrule=0.6pt,
  arc=5pt,
  top=0pt,           
  bottom=0pt,        
  boxsep=2pt,        
  toptitle=2pt,      
  bottomtitle=2pt,   
  listing options={
    language=bash,
    basicstyle=\ttfamily\normalsize,  
    numbers=left,
    numberstyle=\normalsize\color{gray!60},
    numbersep=2pt,
    breaklines=true,
    showstringspaces=false,
    commentstyle=\color{gray!70},
    keywordstyle=\color{black},
    morekeywords={FROM,RUN,WORKDIR,COPY,ENTRYPOINT,ARG,ENV,ADD,CMD,EXPOSE},
  }
}

\newcommand{\my}{\mbox{\emph{Cadre}}\xspace}
\newcommand{\ex}{\mbox{\emph{DodeX}}\xspace}

\newenvironment{takeaway}{
  \par\addvspace{0.3em}  
  \noindent\begin{minipage}{\linewidth}
  \begin{snugshade}
  \setlength{\leftskip}{0.3em}
  \setlength{\rightskip}{0.3em}
  \noindent
}{
  \end{snugshade}
  \end{minipage}
  \par\addvspace{0.3em}  
}

\definecolor{shadecolor}{gray}{0.9}

\graphicspath{{./images/}}

\articletype{RESEARCH ARTICLE - TECHNOLOGY}%

\received{22 May 2026}

\volume{x}
\copyyear{2026}
\startpage{1}
\articledoi{10.1002/0000}

\begin{document}
\title{Taming the Drift: Context-aware Repair of Dockerfile Drift during Software Evolution}
\transtitle{Guidelines for Establishing a Cytometry Laboratory}
\subtranstitle{trans-subtitle}
\author[1,2]{Chengjie Wang}[https://orcid.org/0009-0008-4445-9709]
\author[1,3]{Jingzheng Wu}[https://orcid.org/0000-0001-5561-9829]
\author[1,3]{Xiang Ling}[https://orcid.org/0000-0002-7377-7844]
\author[1]{Tianyue Luo}[https://orcid.org/0000-0001-7407-8255]
\author[1]{Chen Zhao}[https://orcid.org/0009-0005-3386-0335]

\authormark{WANG \textsc{et al.}}
\titlemark{Taming the Drift: Context-aware Repair of Dockerfile Drift during Software Evolution}  

\address[1]{\orgdiv{Intelligent Software Research Center, }\orgname{Institute of Software, Chinese Academy of Sciences, }%
\orgaddress{\state{Beijing, }\country{China}}}

\address[2]{\orgname{University of Chinese Academy of Sciences, }%
\orgaddress{\state{Beijing, }\country{China}}}

\address[3]{\orgdiv{Key Laboratory of System Software (Chinese Academy of Sciences),}
\orgaddress{\state{Beijing, }\country{China}}}

\corres{ Jingzheng Wu (\email{jingzheng08@iscas.ac.cn}) ~|~ Xiang Ling  (\email{lingxiang@iscas.ac.cn})}





\transkeywords{Dockerfile drift | software evolution | automated program repair}

\abstract[ABSTRACT]{Docker has become the de facto standard for reproducible build environments in modern software engineering.
Its benefits are undermined, however, by Dockerfile drift: a divergence between a Dockerfile and its evolving source code that causes builds to fail silently in CI/CD pipelines.
Existing rule-based and retrieval-based repair approaches operate on the Dockerfile in isolation from the surrounding build context and therefore cannot address context-dependent drift failures.
To close this gap, we present \my, a context-aware framework for the automated repair of Dockerfile drift.
The key design insight is that context structure determines repair quality more than context volume.
Specifically, \my performs a static analysis to construct a Context-aware Dependency Graph (CDG) that maps each Dockerfile instruction to its file-level and inter-instruction dependencies.
\my then uses the CDG to guide an LLM in a two-step workflow: the first step selects only the context causally relevant to the observed failure, and the second step generates a targeted patch from that focused context.
We also introduce \ex, a pipeline that continuously mines real-world Dockerfile drift instances from GitHub Actions CI logs and captures the complete build configurations that static-snapshot datasets omit.
Using \ex, we construct $D^3$, a benchmark of 1,040 drift instances each reproducible locally using the exact parameters from the original CI run.
Across $D^3$, \my achieves a repair rate of 35.22\%, which is 2.78$\times$ that of the rule-based baseline and 1.24$\times$ that of the best LLM-based baseline.
The two-step context-selection workflow keeps 95.25\% of prompts below 30k tokens, eliminating the prompt-overflow failures that cause competing LLM-based methods to produce no patch in 41--58 cases per method.
Ablation results confirm that the CDG and the two-step workflow each contribute independently and are mutually reinforcing.
These results indicate that explicit dependency modeling is a durable repair signal: its advantage over diff-only approaches widens as drift ages across commits, suggesting that context-aware IaC maintenance is a productive direction for future
tool development. All the code and dataset are open-sourced at \url{https://github.com/dw763j/Cadre} for review.}

 \transabstract[transABSTRACT]{Docker has become the de facto standard for reproducible build environments in modern software engineering.
Its benefits are undermined, however, by Dockerfile drift: a divergence between a Dockerfile and its evolving source code that causes builds to fail silently in CI/CD pipelines.
Existing rule-based and retrieval-based repair approaches operate on the Dockerfile in isolation from the surrounding build context and therefore cannot address context-dependent drift failures.
To close this gap, we present \my, a context-aware framework for the automated repair of Dockerfile drift.
The key design insight is that context structure determines repair quality more than context volume.
Specifically, \my performs a static analysis to construct a Context-aware Dependency Graph (CDG) that maps each Dockerfile instruction to its file-level and inter-instruction dependencies.
\my then uses the CDG to guide an LLM in a two-step workflow: the first step selects only the context causally relevant to the observed failure, and the second step generates a targeted patch from that focused context.
We also introduce \ex, a pipeline that continuously mines real-world Dockerfile drift instances from GitHub Actions CI logs and captures the complete build configurations that static-snapshot datasets omit.
Using \ex, we construct $D^3$, a benchmark of 1,040 drift instances each reproducible locally using the exact parameters from the original CI run.
Across $D^3$, \my achieves a repair rate of 35.22\%, which is 2.78$\times$ that of the rule-based baseline and 1.24$\times$ that of the best LLM-based baseline.
The two-step context-selection workflow keeps 95.25\% of prompts below 30k tokens, eliminating the prompt-overflow failures that cause competing LLM-based methods to produce no patch in 41--58 cases per method.
These results indicate that explicit dependency modeling is a durable repair signal: its advantage over diff-only approaches widens as drift ages across commits, suggesting that context-aware IaC maintenance is a productive direction for future
tool development. All the code and dataset are open-sourced at \url{https://github.com/dw763j/Cadre} for review.}





\maketitle
\section{Introduction}
\label{sec:intro}
Modern software development is characterized by rapid, iterative evolution, driven by methodologies such as Agile and DevOps~\cite{amaro2022capabilities,el2025systematic}.
Continuous Integration and Continuous Deployment (CI/CD) pipelines are the backbone of this paradigm, requiring consistent and reproducible build environments~\cite{nakarmi2024optimized,soares2022effects}.

Docker has emerged as the de facto standard for achieving environmental consistency across development, testing, and production~\cite{cito2017empirical,docker2020docker,kithulwatta2021adoption}.
Developers encode the environment as a Dockerfile, an executable build script that specifies all dependencies and setup steps from the source repository~\cite{docker2025dockerfile,hilton2016ci}.

In the modern software development process, every source-code commit in a CI/CD pipeline triggers an automated image build and test cycle~\cite{hilton2016ci,vasilescu2015quality,cncf2023survey}.
The health of a Dockerfile, therefore, directly determines the health of the pipeline.
Prior work finds that more than one in four Docker builds in open-source projects fail~\cite{wu2020buildfailures,hilton2017tradeoffs,wu2025towards,wu2020empirical}, and each failure demands days of developer repair effort.
These failures stem from heterogeneous root causes: base-image tag deprecation, network-level package unavailability, and test regressions introduced by dependency updates.
One category is structurally distinct from the others.
\textit{Dockerfile drift} occurs when the Dockerfile fails to absorb a change that has already landed in the application source code, causing the mismatch to persist silently until the next build attempt exposes it.
Drift is an internal co-evolution problem.
Specifically, the fix information resides in the same repository that produced the failure.

Drift manifests as incorrect file paths, mismatched environment variables, changed architecture platforms, and outdated dependency specifications, among other forms.
For instance, a developer may update a dependency version in application configuration files while leaving the corresponding \texttt{ARG} or \texttt{ENV} declarations in the Dockerfile unchanged, breaking the image build.

Dockerfile drift causes significant disruptions in CI/CD pipelines, leading to build failures, deployment delays, and increased maintenance overhead~\cite{ksontini2021refactorings,zhang2018one}.
Repairing drift is non-trivial because Dockerfile instructions are tightly coupled with the build context: the state of the source code, external dependencies, and the target deployment environment at build time.
Three challenges compound the difficulty.

\begin{itemize}
    \item \textbf{C1. Extracting the precise build context from the software:}
    The Docker build process interacts with the source code and external dependencies through a combination of file copies, shell commands, and build-tool invocations.
    Accurately capturing which files and environment variables each instruction depends on requires understanding these interactions, not reading the Dockerfile text alone.
    Existing rule-based methods~\cite{dockle,hadolint,parfum2024empirical} apply static patterns or heuristics and do not model the build context.
    Recent LLM-based approaches~\cite{11025789,11029932,10.1145/3639478.3643083} similarly focus on the Dockerfile and error logs without incorporating the surrounding software context.

    \item \textbf{C2. Inferring dependencies within and across Dockerfile instructions:}
    Instructions can depend on artifacts produced by earlier instructions in non-obvious ways.
    A \texttt{RUN} instruction may operate on files introduced by an earlier \texttt{COPY} step, or reference environment variables set by a prior \texttt{ARG} instruction, without any syntactic marker linking the two.
    These inter-instruction dependencies are not recoverable from any single instruction in isolation and require a global model of the build process.

    \item \textbf{C3. Locating the information relevant to a specific failure:}
    When a build fails, the error log identifies the failing instruction but rarely points directly to the root cause, especially when the failure originates from a context change introduced in a prior commit.
    A systematic approach is needed to trace back from the observed error to the specific source-code changes that caused the drift, without requiring human inspection.
\end{itemize}

The key insight behind our solution is that \emph{context structure determines repair quality more than context volume}: knowing which files each instruction depends on, and how those dependencies propagate across instruction boundaries, is more valuable than supplying the LLM with a large unstructured context.
Building on this insight, we propose \textbf{\my} (\underline{C}ontext-\underline{a}ware \underline{d}rift \underline{re}pair framework), which models the build context explicitly and uses LLM-based reasoning to generate targeted repairs.
\my addresses each challenge with a dedicated component.

\textbf{The \textit{Instruction-level Context Profiler}} addresses C1 by makeing a stateful simulation of the Docker build process, tracking the working directory, environment variables, and files in the container at each instruction.
Rather than treating build-tool invocations as opaque shell commands, it applies build system parsers to infer the file dependencies that each invocation implies.

\textbf{The \textit{Context-aware Dependency Graph}} (CDG) addresses C2 by converting the per-instruction context profile into a directed graph that makes inter-instruction dependencies explicit.
File vertices connect to the instructions that depend on them, and sequential edges represent the control flow of the build, allowing a repair agent to trace failure causes across instruction boundaries.

\textbf{The \textit{Context-refined Repair}} workflow addresses C3 by using the CDG to guide a two-step LLM interaction.
Rather than supplying all available context in a single prompt, the first step asks the LLM to select the most relevant files from the CDG and the code-change diff; the second step retrieves those files and generates a targeted repair.
This separation keeps each prompt focused and within the model's operational context limit.

Progress in this area has also been hampered by the absence of realistic benchmarks.
Existing Dockerfile repair datasets~\cite{9402688,11025789,11029932} capture the Dockerfile text but not the build configuration or dynamic build parameters required to reproduce builds locally, making evaluation results difficult to interpret.
We therefore design \textbf{\ex} (\underline{Do}ckerfile \underline{d}rift e\underline{X}tractor), a pipeline that continuously mines Dockerfile drift instances from GitHub Actions CI logs and captures their complete build configurations, including dynamic build arguments and target platform specifications.
Using \ex, we construct $D^3$, the first Dockerfile drift benchmark in which each instance is locally reproducible using the exact parameters recorded from the original CI run.

We evaluate \my on $D^3$, comparing it against a rule-based baseline~\cite{parfum2024empirical}, a retrieval-based baseline~\cite{11029932}, and a vanilla LLM baseline.
The context-aware dependency modeling substantially raises the repair rate.
\my achieves a repair rate of 35.22\% on the benchmark, against 9.34\% for the rule-based baseline and 28.48\% for the best LLM-based baseline without context modeling.
Beyond accuracy, context selection eliminates prompt-overflow failures entirely.
Each LLM-based baseline produces between 41 and 58 cases per method in which no patch is generated, because the prompt exceeds the model's context limit.
\my produces zero such cases: its two-step workflow filters context before loading file content, keeping every prompt within the operational range.
The performance advantage widens further as drift becomes stale.
When a build failure persists across multiple commits, the most recent code diff becomes an increasingly unreliable repair signal.
At five or more commits of distance, \my's repair rate of 25.9\% exceeds the best baseline's 19.7\% by a proportionally larger margin than at the first commit, confirming that the CDG provides a signal that is robust to commit distance.

\begin{figure*}[b]
\begin{dockerlisting}{\textbf{Listing 1}: Dockerfile of \texttt{httpx} at the failing commit bb3154ff~\cite{httpx}}
# Stage 1: Build
FROM golang:1.21.4-alpine AS builder
RUN apk add --no-cache git build-base gcc musl-dev
WORKDIR /app
COPY . /app
RUN go mod download
RUN go build ./cmd/httpx
# Stage 2: Runtime
FROM alpine:3.18.2
RUN apk upgrade --no-cache && apk add --no-cache bind-tools ca-certificates chromium
COPY --from=builder /app/httpx /usr/local/bin/
ENTRYPOINT ["httpx"]
\end{dockerlisting}
\end{figure*}

In summary, the contributions of this paper are as follows.

\begin{itemize}
    \item \textbf{\my:}
    A context-aware Dockerfile drift repair framework grounded on the insight that context structure determines repair quality more than context volume.
    \my realizes this through an Instruction-level Context Profiler, a Context-aware Dependency Graph, and a two-step LLM repair workflow that together relieve the prompt-overflow failures that affect all competing methods while achieving a repair rate 2.78$\times$ that of the rule-based baseline and 1.24$\times$ that of the best LLM-based baseline.

    \item \textbf{\ex and $D^3$:}
    A continuously operable pipeline that mines Dockerfile drift instances from GitHub Actions CI logs and captures the complete build configuration, \textit{i.e.}, including dynamic build arguments and multi-platform specifications, that static-snapshot datasets omit, together with $D^3$, the first Dockerfile drift benchmark in which every instance is locally reproducible using the exact parameters from the original CI run.

    \item \textbf{Empirical evaluation:}
    A comparison of \my against rule-based, retrieval-based, and vanilla LLM baselines under a reproducible evaluation protocol, covering repair effectiveness, token efficiency, robustness to stale drift, and component-level ablation, providing a multi-dimensional characterization of context-aware repair.
\end{itemize}

We organize the remainder of the paper as:
Section~\S\ref{sec:background} provides background on Dockerfiles and the Docker build process.
Section~\S\ref{sec:methodology} details the design of \my.
Section~\S\ref{sec:experiment} describes the experimental setup and the construction of $D^3$.
Section~\S\ref{sec:results} presents experimental results.
Section~\S\ref{sec:discussion} discusses findings and threats to validity.
Section~\S\ref{sec:related_work} reviews related work.
Section~\S\ref{sec:conclusion} concludes.

\section{Background}
\label{sec:background}
 
To understand why Dockerfile drift is difficult to repair automatically, this section covers two prerequisites: the build-time semantics of Dockerfile instructions, and the precise technical meaning of Dockerfile drift as an evolutionary phenomenon.
Both are illustrated with a real drift instance from the $D^3$ dataset.
 
\subsection{Docker Build Process and Context Coupling}
\label{sec:bg_docker}
 
A Dockerfile is an ordered sequence of instructions that Docker executes to assemble a container image layer by layer~\cite{docker2025dockerfile}.
Understanding how each instruction type interacts with the build environment is necessary to understand why a repair tool cannot operate on the Dockerfile text in isolation.
 
\textbf{State-accumulating instructions.}
Three instruction types modify the build state without producing file content.
\texttt{FROM} designates the base image for a build stage and implicitly establishes the runtime environment, including the version of any language toolchain bundled in that image.
\texttt{WORKDIR} sets the working directory inside the container; all subsequent relative paths in \texttt{COPY}, \texttt{ADD}, and \texttt{RUN} instructions are resolved against this value.
\texttt{ARG} and \texttt{ENV} declare build-time and runtime variables, respectively; later instructions can reference these variables via \texttt{\$\{VAR\}} parameter.
 
\textbf{Context-ingesting instructions.}
\texttt{COPY} and \texttt{ADD} transfer files from the host build context, which is typically the repository root, into the container image.
Their source paths are resolved against the repository root, making them a direct coupling point between the Dockerfile and the source tree.
 
\textbf{Execution instructions.}
\texttt{RUN} executes a shell command inside the container at build time.
Its behavior depends on three inherited states: the working directory established by prior \texttt{WORKDIR} instructions, the environment variables accumulated by prior \texttt{ARG} and \texttt{ENV} instructions, and the files present in the container from prior \texttt{COPY} or \texttt{ADD} instructions.
Critically, none of these dependencies is syntactically visible in the \texttt{RUN} instruction itself.
 
\textbf{Multi-stage builds.}
A Dockerfile may contain multiple \texttt{FROM} instructions, each initiating a new build stage with its own independent working directory, environment, and file set.
Files may be transferred from an earlier stage to a later one via \texttt{COPY --from=\textit{stage}}, creating cross-stage dependencies that are not visible from the receiving instruction alone.

\textbf{Listing~1} shows a representative two-stage Dockerfile from the httpx project~\cite{httpx}.
The first stage at lines~1--8 compiles the Go binary: \texttt{WORKDIR} on line~4 establishes \texttt{/app} as the working directory; \texttt{COPY} on line~5 brings the entire repository into \texttt{/app}, including \texttt{go.mod} and \texttt{go.sum}; \texttt{RUN go mod download} on line~6 then resolves Go module dependencies against those files.
The second stage at lines~10--14 assembles the runtime image and retrieves the compiled binary from the first stage via \texttt{COPY --from=builder} on line~13, a cross-stage dependency.

\subsection{Dockerfile Drift}
\label{sec:bg_drift}
 
Dockerfile drift occurs when a code change in the source repository invalidates one or more Dockerfile instructions, causing the next CI build to fail.
The defining characteristic of drift is that the Dockerfile itself is syntactically unchanged; the failure arises from the broken correspondence between Dockerfile semantics and the evolved repository state.
 
The httpx instance in Listing~1 illustrates drift concretely.
In this commit, the developers updated \texttt{go.mod} to declare \texttt{go 1.23.0} as the minimum required toolchain, while also bumping dozens of transitive dependency versions.
The Dockerfile was not modified.
When GitHub Actions executed the Docker build, the \texttt{RUN go mod download} instruction at line~6 failed with the following error: \textit{go: go.mod requires go >= 1.23.0 (running go 1.21.4; GOTOOLCHAIN=local).}
 
The error message correctly attributes the failure to line~6, but the root cause is not located there.
Line~6 fails because two earlier instructions have produced an incompatible state: \texttt{FROM golang:1.21.4-alpine} on line~2 establishes a Go 1.21.4 runtime, and \texttt{COPY . /app} on line~5 brings in the updated \texttt{go.mod} that declares a minimum version of Go 1.23.0.
No static analysis of any single instruction reveals this incompatibility.
A repair tool must trace the dependency path from the \texttt{FROM} instruction through the \texttt{COPY} instruction to the \texttt{RUN} instruction to identify that the correct fix is to update the base image on line~2.
 
This gap between the error's visible location and its actual root cause is the central challenge that motivates \my's design.
Without a model of how the build-time state propagates across instructions, a repair tool cannot reliably distinguish the instruction that needs to change from the instruction that happens to fail.

\section{Methodology}
\label{sec:methodology}

\subsection{Overview}

\begin{figure*}[t]
  \centering
  \includegraphics[width=0.9\linewidth]{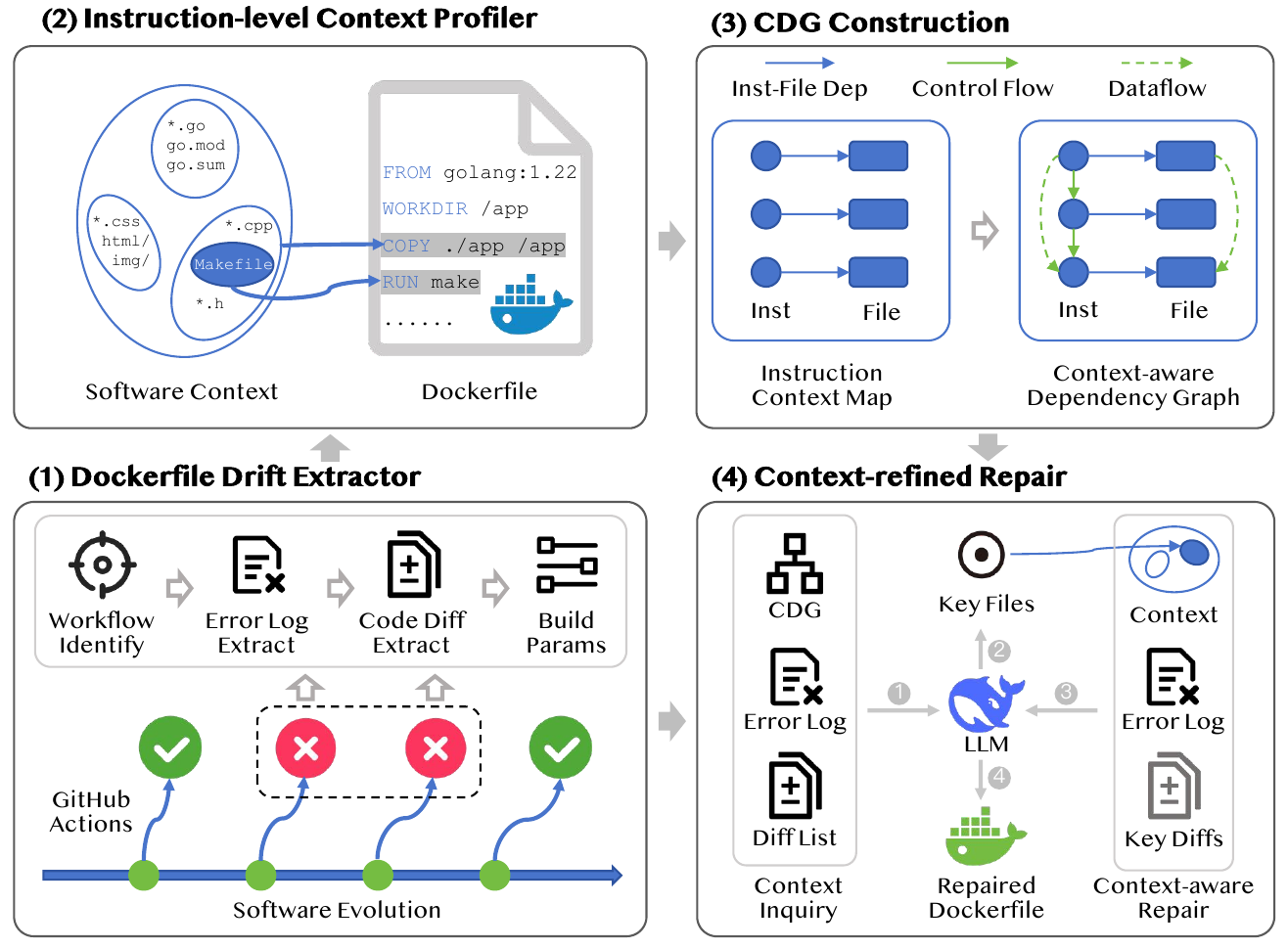}
  \caption{The overview of the \my framework.}
  \label{fig:overview}
\end{figure*}

This section details the method we designed to repair the Dockerfile Drift problem by dealing the  three challenges identified in Section~\ref{sec:intro}, \textit{i.e.}, extracting build-time context (C1), modeling cross-instruction dependencies (C2), and locating causally relevant information (C3).
\my deals these challenges with dedicated components, organized into four phases together with the \ex extractor, as illustrated in Figure~\ref{fig:overview}.

\textbf{(1) Dockerfile Drift Extractor (\ex):}
The first step in repairing Dockerfile drift is locating instances of it in the wild.
Existing Dockerfile repair datasets~\cite{9402688,11025789,11029932} focus on the Dockerfile text alone, without capturing the build context that surrounds it.
\ex uses GitHub Actions as a natural observation point that simultaneously records software evolution and Docker build outcomes, enabling the extraction of drift instances together with their full build configurations and dynamic build parameters.

\textbf{(2) Instruction-level Context Profiler:}
This phase performs an instruction-level static analysis to profile the build context, maintaining a context state that tracks file accesses, working-directory changes, and environment variables.
By processing each Dockerfile instruction against this evolving state, the profiler produces a detailed context profile for the entire Dockerfile to deal with C1.

\textbf{(3) Context-aware Dependency Graph (CDG) Construction:}
This phase models the dependencies among Dockerfile instructions as a directed acyclic graph, using the context profile produced in the previous phase.
The graph makes explicit both the control flow among instructions and the data-flow relationships between instructions and the repository files they access, enabling systematic reasoning about failure propagation paths that can deal with C2.

\textbf{(4) Context-refined Repair:}
This phase employs an LLM to reason about which context is essential for repairing the specific build failure, guided by the CDG and the failure log.
Through a two-step agentic workflow, the LLM first selects the most relevant files and then generates a targeted repair to deal with C3.

We detail each phase in the following sections.

\subsection{Dockerfile Drift Extractor}
\label{sec:setup_extractor}

Existing studies on Dockerfile analysis primarily rely on static snapshots of Dockerfiles collected at a single point in time~\cite{9402688,11029932}.
Such datasets capture syntactical issues but cannot represent Dockerfile drift, which is an evolutionary problem: it arises from desynchronization between a project's source code and its build configuration across successive commits.
To fill this gap, we developed the \underline{Do}ckerfile \underline{d}rift e\underline{X}tractor (\textbf{\ex}), a systematic pipeline that mines genuine drift instances directly from the software evolution history of real-world projects.

The core design rationale of \ex is to use GitHub Actions as a natural laboratory for observing Dockerfile drift.
A GitHub Actions workflow is triggered by a \texttt{push}, \texttt{pull\_request}, or other commit-level event that represents a discrete software change.
When a Docker build step within such a workflow fails, the failure provides a verifiable signal that the current source code has diverged from the Dockerfile.
\ex captures this signal together with the complete build configuration, making each extracted instance fully reproducible.

The \ex pipeline automates the discovery and capture of these instances through the following stages.

\textbf{(1) Repository Collection:}
The pipeline begins by collecting a large set of high-quality open-source projects from GitHub.
\ex uses the GitHub GraphQL API and applies filters on repository creation date, primary programming language, star count, and the presence of a \texttt{Dockerfile} to select projects that are mature and actively maintained.

\textbf{(2) Workflow Identification:}
\ex downloads and parses the workflow YAML files from the collected repositories.
It selects only those workflows that contain an explicit Docker build step, such as the widely used \texttt{docker/build-push-action} action, and discards CI jobs unrelated to containerized builds.

\textbf{(3) Failure Log Acquisition:}
For each Docker-centric workflow, \ex uses the GitHub REST API to fetch its complete execution history.
It targets all failed workflow runs and downloads their line-by-line execution logs for offline analysis.

\textbf{(4) Drift Instance Extraction and Context Capture:}
The downloaded logs are parsed to isolate failures that occurred specifically within the Docker build step.
\ex applies a two-stage filtering procedure to distinguish Docker build failures from other CI failures in the same workflow run.
Firstly, \ex identifies the workflow step that triggered the failure by matching the step name and action type against a curated list of Docker-build action identifiers, \textit{e.g.}, \texttt{docker/build-push-action} and \texttt{docker build} commands.
Secondly, \ex verifies that the failure log contains Docker-specific error patterns, such as layer-build errors and Dockerfile instruction traces, rather than generic exit codes produced by testing or deployment steps.
Only instances that pass both stages are retained as confirmed drift instances.

For each confirmed instance, \ex extracts and stores the following metadata:

\begin{itemize}
    \item The full build error log from the failed Docker build step.
    \item The complete Dockerfile build configuration, including the target Dockerfile path, build context path, dynamic build arguments, and target platforms.
          This information is extracted from both the workflow file at the failing commit and the build log itself; \ex cross-validates the two sources to resolve discrepancies and obtain accurate parameter values.
    \item The commit hash of the failed build, which allows the exact source repository state that triggered the failure to be reconstructed locally.
    \item The code-change diffs introduced by the failing commit, which provide the evolutionary signal used by repair tools.
\end{itemize}

The output of the \ex pipeline is a collection of drift instances.
Each instance encapsulates not only a failing Dockerfile and an error log, but the entire evolutionary context: the specific software change that triggered the failure and the precise, dynamic build parameters recorded by the CI environment.
This provides the foundation for constructing our high-fidelity dataset.

\subsection{Instruction-level Context Profiler}
\label{sec:method_profile}

\begin{algorithm}[t!]
\caption{Instruction-level Context Profiler}
\label{alg:context_profiling}
\begin{algorithmic}[1]
\State \textbf{Input:} $D$: Parsed Dockerfile Instructions, $R$: Repository root path
\State \textbf{Output:} $ICM$: Instruction Context Map
\Function{Profiling}{$D, R$}
    \State $StageStates \gets \text{new map of stage names to } \sigma$
    \State $ICM \gets \text{new map of instructions to } \sigma$
    \State $current\_stage \gets \text{null}$

    \For{each instruction $I$ in $D$}
        \State $\sigma \gets StageStates[current\_stage]$

        \If{$I.type = \texttt{FROM}$}
            \State $current\_stage \gets I.stage\_name$
            \State $\sigma_{new} \gets (\delta \gets \text{``/''}, \varepsilon \gets \bot, \Phi \gets \emptyset)$
            \State $StageStates[current\_stage] \gets \sigma_{new}$
            \State $\sigma \gets \sigma_{new}$

        \ElsIf{$I.type \in \{\texttt{WORKDIR}, \texttt{ARG}, \texttt{ENV}\}$}
            \State $\sigma \gets \text{UpdateState}(\sigma, I)$

        \ElsIf{$I.type \in \{\texttt{COPY}, \texttt{ADD}\}$}
            \If{$I \text{ has } \texttt{--from} \text{ flag}$}
                \State $source\_stage \gets I.source\_stage\_name$
                \State $\sigma_{src} \gets StageStates[source\_stage]$
                \State $SourceFiles \gets \text{FindFiles}(\sigma_{src}.\Phi, I.src)$
            \Else
                \State $SourceFiles \gets \text{ResolveHostPaths}(R, I.src)$
            \EndIf
            \State $\sigma.\Phi \gets \sigma.\Phi \cup \text{MapToDest}(\text{SourceFiles}, I.dest)$

        \ElsIf{$I.type \in \{\texttt{RUN}, \texttt{SHELL}, \texttt{CMD}\}$}
            \State $commands \gets \text{Preprocess}(\text{I.value}, \sigma.\varepsilon)$
            \For{each $cmd$ in $commands$}
                \State $\sigma \gets \text{UpdateStateForPathChanges}(\sigma, cmd)$
                \State $inferred\_files \gets \text{AnalyzeBuild}(cmd, \sigma)$
                \State $\sigma.\Phi \gets \sigma.\Phi \cup inferred\_files$
            \EndFor
        \EndIf
        \State $ICM[I] \gets \text{copy}(\sigma)$
    \EndFor
    \State \textbf{return} $ICM$
\EndFunction
\end{algorithmic}
\end{algorithm}

C1 requires that every Dockerfile instruction be associated with the precise build-time state visible to it at the moment of execution.
This state is determined by the cumulative effect of all prior instructions in the same build stage.
Static analysis of the Dockerfile text alone is therefore insufficient: the working directory, active environment variables, and the set of available files are all functions of prior execution, not of text structure.
Correct context extraction requires simulating the build process instruction by instruction, propagating a shared state model across the full instruction sequence.

The Instruction-level Context Profiler implements this simulation.
It traverses the Dockerfile's instructions in order and maintains a model of the evolving container environment.
By the time a repair agent queries the profiler for the context of a failed instruction, the profiler has already recorded the full state that instruction could see at build time.

\textbf{Context State.}
The core data structure of this simulation is the \textit{Context State}, which formally captures the container's environment at a given point in the build.
We define the Context State immediately before instruction $I_j$ executes as:
\begin{equation}
\label{eqa:context_state}
\sigma_j = \bigl(\,\delta_j,\;\varepsilon_j,\;\Phi_j\,\bigr)
\end{equation}
where $\delta_j \in \mathcal{P}$ is the current working directory, drawn from the set of all absolute container paths $\mathcal{P}$; $\varepsilon_j : \mathcal{K} \rightharpoonup \mathcal{V}$ is a partial function mapping environment variable names $\mathcal{K}$ to their resolved string values $\mathcal{V}$, populated by \texttt{ENV} and \texttt{ARG} instructions; and $\Phi_j \subseteq \mathcal{F}$ is the set of file paths present in the current build stage immediately before $I_j$ runs, where $\mathcal{F}$ denotes the universe of all file paths.
The profiler initializes each build stage with $\sigma_0 = (\texttt{/},\;\bot,\;\emptyset)$, where $\bot$ denotes the empty mapping, and advances the state through a deterministic transition function:
\begin{equation}
\label{eq:state_transition}
\sigma_{j+1} = \mathcal{T}(I_{j+1},\;\sigma_j)
\end{equation}
The concrete semantics of $\mathcal{T}$ are instruction-type-specific and defined in the per-type analysis below.
A key invariant of this simulation is that $\Phi$ is monotonically non-decreasing within a build stage: $\Phi_j \subseteq \Phi_{j+1}$ for all $j$.
This reflects Docker's union filesystem, in which each layer accumulates files from prior layers and never removes them within a stage.

Modern Dockerfiles frequently employ multi-stage builds to separate build-time from runtime dependencies.
\my respects this structure by instantiating an independent Context State $\sigma$ for each build stage initiated by a \texttt{FROM} instruction.
A map of these stage-specific states is maintained throughout the analysis, enabling correct resolution of cross-stage file transfers, \textit{e.g.}, \texttt{COPY --from=<stage>}.
We provide the profiling pseudocode in Algorithm~\ref{alg:context_profiling} and describe the per-instruction-type analysis below.

\textbf{(1) State-Modifying Instructions:}
Directives such as \texttt{WORKDIR}, \texttt{ARG}, and \texttt{ENV} directly update the Context State.
A \texttt{WORKDIR} instruction updates $\delta$, resolving relative paths against the current value.
\texttt{ARG} and \texttt{ENV} instructions populate $\varepsilon$, with later definitions overriding earlier ones as per Docker's semantics.

\textbf{(2) Context-Ingesting Instructions:}
\texttt{COPY} and \texttt{ADD} instructions ingest files into the container.
The profiler resolves their source paths against either the host repository root or a previous stage's file set $\Phi$, and adds the corresponding file paths to $\Phi$ of the current state.
The destination path is computed relative to the current working directory $\delta$.
This step explicitly links each Dockerfile instruction to its file-level dependencies.

\textbf{(3) Execution Instructions:}
Analyzing \texttt{RUN}, \texttt{CMD}, and \texttt{ENTRYPOINT} instructions requires handling arbitrary shell commands.
For a \texttt{RUN} instruction, the profiler proceeds as below.

\begin{itemize}
    \item \textbf{Preprocessing:} Chained shell commands, \textit{i.e.}, joined by \texttt{\&\&} or \texttt{;}, are decomposed into a sequence of commands.
    
    \item \textbf{Variable substitution:} Environment variable references, \textit{e.g.}, \texttt{\$\{VAR\}}, in each command are substituted using the current environment map $\varepsilon$.
    
    \item \textbf{Build-system-aware analysis:} Rather than treating commands as opaque strings, the profiler applies build-tool-specific parsers to infer file dependencies from recognized build-tool invocations.
    For example, a \texttt{go build} command triggers the Go parser, which infers dependencies on \texttt{go.mod}, \texttt{go.sum}, and all \texttt{*.go} source files within the working directory $\delta$.
    A \texttt{uv sync} command triggers the Python-uv parser, which infers a dependency on \texttt{pyproject.toml}.
    This build-system-aware approach allows the profiler to recover file dependencies that are implicit in the build-tool contract rather than stated in the Dockerfile text.
\end{itemize}

The profiler implements build-system-aware parsers for 11 major ecosystems.
Table~\ref{tab:build_parsers} summarizes the trigger commands and inferred file dependencies for each ecosystem.

\begin{table*}[t]
\centering
\caption{Build-system-aware parsers implemented in \my. Each parser is triggered by a recognized command prefix and infers the corresponding file dependencies relative to the current working directory $\delta$.}
\label{tab:build_parsers}
\resizebox{\linewidth}{!}{%
\begin{tabular}{llp{5.2cm}}
\toprule
\textbf{Ecosystem} & \textbf{Trigger commands} & \textbf{Inferred file dependencies} \\
\midrule
Go        & \texttt{go build}, \texttt{go test}, \texttt{go install},
            \texttt{go run}, \texttt{go mod}, \texttt{go get},
            \texttt{go fmt}, \texttt{go vet}
          & \texttt{go.mod}, \texttt{go.sum}, \texttt{*.go} in $\delta$ \\
npm       & \texttt{npm install}, \texttt{npm ci}, \texttt{npm build},
            \texttt{npm run}, \texttt{npm start}, \texttt{npm test}
          & \texttt{package.json}, \texttt{package-lock.json} \\
yarn      & \texttt{yarn install}, \texttt{yarn build},
            \texttt{yarn run}, \texttt{yarn start}, \texttt{yarn test}
          & \texttt{package.json}, \texttt{yarn.lock} \\
pnpm      & \texttt{pnpm install}, \texttt{pnpm run},
            \texttt{pnpm build}, \texttt{pnpm add}
          & \texttt{package.json}, \texttt{pnpm-lock.yaml} \\
pip       & \texttt{pip install}, \texttt{python install},
            \texttt{poetry install}, \texttt{pipenv install}
          & \texttt{requirements*.txt}, \texttt{setup.py},
            \texttt{pyproject.toml}, \texttt{Pipfile} \\
uv        & \texttt{uv sync}, \texttt{uv run}, \texttt{uv lock},
            \texttt{uv add}, \texttt{uv build}
          & \texttt{pyproject.toml}, \texttt{uv.lock} \\
Maven     & \texttt{mvn}, \texttt{mvnw}
          & \texttt{pom.xml} \\
Gradle    & \texttt{gradle}, \texttt{gradlew}
          & \texttt{build.gradle}, \texttt{build.gradle.kts},
            \texttt{settings.gradle}, \texttt{settings.gradle.kts} \\
Cargo     & \texttt{cargo build}, \texttt{cargo test}, \texttt{cargo run},
            \texttt{cargo check}, \texttt{cargo clippy},
            \texttt{cargo install}
          & \texttt{Cargo.toml}, \texttt{Cargo.lock} \\
.NET      & \texttt{dotnet build}, \texttt{dotnet restore},
            \texttt{msbuild}, \texttt{nuget}
          & \texttt{*.csproj}, \texttt{*.vbproj}, \texttt{*.fsproj},
            \texttt{*.sln} \\
Composer  & \texttt{composer install}, \texttt{composer update},
            \texttt{composer require}
          & \texttt{composer.json}, \texttt{composer.lock} \\
Make      & \texttt{make}
          & \texttt{GNUmakefile}, \texttt{Makefile},
            \texttt{*.c}, \texttt{*.cpp}, \texttt{*.h} in $\delta$ \\
CMake     & \texttt{cmake --build}, \texttt{cmake -D...}
          & \texttt{CMakeLists.txt} \\
\bottomrule
\end{tabular}%
}
\end{table*}

The parser system is implemented as a pluggable architecture.
A dispatcher module routes each command to the appropriate parser based on the command's prefix.
A new parser can add new build systems without modifying existing ones.

The final output of the Instruction-level Context Profiler is the \textit{Instruction Context Map (ICM)}: a mapping from each instruction $I$ in the Dockerfile to the Context State $\sigma$ that existed immediately before its execution.
The ICM provides the foundational data for constructing the dependency graph in the next phase.

\subsection{Context-aware Dependency Graph Construct}
\label{sec:method_cdg}

\begin{figure}[b]
  \centering
  \includegraphics[width=\columnwidth]{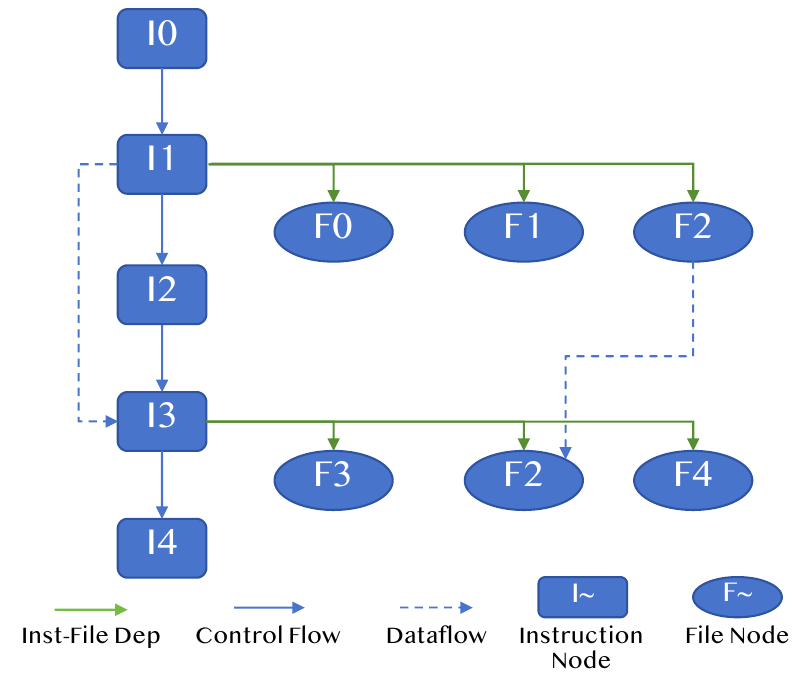}
  \caption{Illustration of the Context-aware Dependency Graph.}
  \label{fig:cdg}
\end{figure}

C2 requires a representation of how dependencies propagate across instruction boundaries.
A per-instruction context profile is insufficient for this purpose.
Consider a \texttt{COPY} instruction that introduces a file later consumed by a \texttt{RUN} instruction.
When the build fails at the \texttt{RUN} step, the root cause is a change to the file introduced by \texttt{COPY}.
This causal link spans an instruction boundary and is invisible when instructions are analyzed in isolation.
The repair agent must be able to trace it.
To model these cross-instruction relationships explicitly, we construct the Context-aware Dependency Graph (CDG), a directed representation of both the control flow and data flow of the build process.
Formally, the CDG is a directed acyclic graph $G = (V, E)$ as below.

\begin{algorithm}[b]
\caption{CDG Construction}
\label{alg:cdg_construction}
\begin{algorithmic}[1]
\State \textbf{Input:} $D$: Parsed Dockerfile Instructions, $ICM$: Instruction Context Map
\State \textbf{Output:} $G = (V, E)$: Context-aware~Dependency~Graph
\Function{ConstructCDG}{$D, ICM$}
    \State $V_I \gets \emptyset$, $V_F \gets \emptyset$, $E_S \gets \emptyset$, $E_D \gets \emptyset$
    \State $v_{prev} \gets \text{null}$

    \For{each instruction $I_j$ in $D$}
        \State $v_j \gets \text{create\_instruction\_vertex}(I_j)$
        \State $V_I \gets V_I \cup \{v_j\}$
        \If{$v_{prev} \neq \text{null}$}
            \State $E_S \gets E_S \cup \{(v_{prev}, v_j)\}$
        \EndIf
        \State $v_{prev} \gets v_j$
    \EndFor

    \For{each instruction $I_j$ in $D$}
        \State $\sigma_j \gets ICM[I_j]$
        \State $v_j \gets \text{get\_vertex}(I_j)$
        \For{each file path $F_k$ in $\sigma_j.\Phi$}
            \State $v_k \gets \text{get\_or\_create\_file\_vertex}(F_k)$
            \State $V_F \gets V_F \cup \{v_k\}$
            \State $E_D \gets E_D \cup \{(v_k, v_j)\}$
        \EndFor
    \EndFor

    \State $V \gets V_I \cup V_F$, $E \gets E_S \cup E_D$
    \State \textbf{return} $G = (V, E)$
\EndFunction
\end{algorithmic}
\end{algorithm}

\begin{itemize}
    \itemsep0em
    \item \textbf{Vertices ($V$):} $V = V_I \cup V_F$, where the two sets are disjoint.
    \begin{itemize}
    \itemsep0em
        \item $V_I$: instruction vertices, one per Dockerfile instruction.
        \item $V_F$: file vertices, one per unique file path from the build context that is referenced by at least one instruction.
              When multiple instructions depend on the same file path, a single shared $V_F$ is used.
              The CDG thus does not duplicate structural information across shared dependencies.
    \end{itemize}
    \item \textbf{Edges ($E$):} $E = E_S \cup E_D$, where the two sets are disjoint.
    \begin{itemize}
    \itemsep0em
        \item $E_S$: sequential edges.
              For any two consecutively appearing instructions $I_j$ and $I_{j+1}$, a sequential edge $(v_j, v_{j+1}) \in E_S$ represents the control flow of the build.
        \item $E_D$: dataflow edges.
              If instruction $I_j$ depends on file $F_k$ (by its file set $\Phi_j$), a dataflow edge $(v_k, v_j) \in E_D$ represents the ``read-from'' dependency from context to instruction.
    \end{itemize}
\end{itemize}

\textbf{Transitive dependencies.}
The CDG represents only direct dependencies.
Specifically, a dataflow edge $(v_k, v_j)$ exists if and only if instruction $I_j$ directly references file $F_k$ according to the ICM.
Transitive relationships are not added as explicit edges.
For example, a change to $F_k$ may affect instruction $I_m$ via an intermediate instruction $I_j$; adding this as a direct edge would cause quadratic graph growth.
Instead, such relationships are recoverable by path traversal and are exploited as such during the repair phase.
This design keeps the graph compact while preserving all dependency information needed for reasoning.

\begin{table*}[b]
\centering
\caption{Comparison of $D^3$ with existing Dockerfile-related datasets.}
\label{tab:dataset_comparison}
\begin{tabular}{lccc}
\toprule
\textbf{Characteristic} & \textbf{$D^3$} & \textbf{Shipwright~\cite{9402688}} & \textbf{Ksontini. et al\cite{11025789}} \\
\midrule
Source of Data            & Real-world Evolution & GitHub Snapshots & GitHub Snapshots \\
Includes Full Build Context & Yes & No & No \\
Includes Build Arguments  & Yes & No & No \\
Includes Build Architectures & Yes & No & No \\
\bottomrule
\end{tabular}
\end{table*}

The construction procedure is detailed in Algorithm~\ref{alg:cdg_construction}.
All instruction vertices are created first and linked by sequential edges.
The algorithm then iterates over the ICM: for each instruction, it examines the associated file set $\Phi$, creates a file vertex for each unique path (or reuses an existing one), and adds a dataflow edge from the file vertex to the instruction vertex.

The resulting CDG supports efficient causal reasoning about drift root causes.
Given that a code change modifies file $F_k$ and the build fails at instruction $I_j$, a repair agent traverses the CDG to determine whether a dependency path from $v_k$ to $v_j$ exists.
If such a path exists, the code change is a plausible cause of the observed failure.
The CDG thus supports the context-refined repair phase.

\subsection{Context-refined Repair}
\label{sec:method_repair}

C3 requires identifying the minimal context that is causally sufficient for generating a correct repair.
Supplying all available context to the LLM is counterproductive.
The token budget is consumed by files irrelevant to the specific failure, and repair accuracy degrades as the model's attention spreads across noise.
The repair phase must therefore identify, from the CDG and the failing commit's change set, the smallest subset of context that is sufficient to generate a correct repair.

To achieve this, we propose Context-refined Repair, which uses the reasoning capability of an LLM to actively select the context it needs.
Rather than a single-shot repair attempt, we design a two-step agentic workflow: \textbf{(1) Context Inquiry} and \textbf{(2) Context Augmentation and Repair Generation}.
This design follows the retrieval-augmented generation paradigm~\cite{lewis2020rag}, in which a retriever supplies evidence that the generator alone cannot reliably hallucinate~\cite{chen2021codex}.

\textbf{(1) Context Inquiry.}
The goal of this step is to identify which files and code regions are most relevant to the specific build failure, before any file contents are loaded.

The workflow begins by assembling an initial context package $\mathcal{C}_{init}$ containing three inputs: the build failure log $L_{err}$, the list of changed files $\Delta_F$ derived from the failing commit's diff, and the CDG $G$.
To bound the size of the serialized graph, a pruning strategy is applied.
For any instruction vertex whose in-degree from file vertices exceeds a threshold $\alpha = 20$, its file dependencies are condensed into directory-level representations.
This produces a pruned graph $G'$ that preserves the high-level dependency structure while eliminating noise from instructions like ``\texttt{COPY . .}'' that reference entire directory trees.

\textbf{Prompt structure.}
The inquiry prompt $P_{inquiry}$ is organized into five sections: (1) the serialized CDG (\texttt{BUILD CHANNELS}), listing each instruction alongside its linked prior instructions and associated file dependencies; (2) the build failure log; (3) a structured summary of code changes, containing new, modified, and deleted file lists; (4) a system role description; and (5) an output specification requiring a JSON object with fields for new, modified, and deleted files selected from $\Delta_F$, and a \texttt{key\_files} list specifying file paths and grep keywords for non-diff context files.
The formal representation is:

\begin{equation}
    R_{files} = \text{LLM}(P_{inquiry}(L_{err}, \Delta_F, G'))
\end{equation}

where $R_{files}$ is the LLM's structured response identifying the files and grep keywords needed for repair.
For example, given an \texttt{npm install} failure, the LLM would typically request \texttt{package.json} and \texttt{package-lock.json} from the changed file list, along with any npm-related grep keywords.

\textbf{(2) Context Augmentation and Repair Generation.}
The second step retrieves the specific context identified by the LLM and uses it to generate the repair.

For each file in $R_{files}$ that belongs to the changed set $\Delta_F$, the full diff content is retrieved from the repository.
For each file in $R_{files}$ that belongs to the broader build context, the ICM maps the path back to the repository, the file's content is then retrieved, and it is filtered using the grep keywords supplied by the LLM.
This produces the augmented context $\mathcal{C}_{aug}$, which contains only the information the LLM identified as relevant.

\textbf{Prompt structure.}
The repair prompt $P_{repair}$ is organized into five sections: (1) the original buggy Dockerfile; (2) the serialized CDG augmented with the retrieved key-file contents (\texttt{BUILD COMMAND} section); (3) the detailed code changes for the selected diff files; (4) the build failure log; and (5) an output specification requiring only a fenced Dockerfile block with no additional commentary.
The formal representation is:

\begin{equation}
    P_{patch} = \text{LLM}(P_{repair}(L_{err}, \mathcal{C}_{aug}, D_{orig}))
\end{equation}

where $D_{orig}$ is the original failing Dockerfile and $P_{patch}$ is the generated patch.

\textbf{Design rationale.}
The two-step workflow separates the \emph{what-to-look-at} decision from the \emph{how-to-fix} decision, delegating each to a different LLM capability.
The first step exploits the LLM's ability to reason about failure semantics from a compact structural description.
The second step then applies its code-generation capability to a focused, task-relevant context.
This separation avoids the token waste that a single-prompt approach incurs by conflating context selection with patch generation.
It also grounds the repair in verifiable evidence from the CDG rather than in the LLM's parametric memory alone.

\section{Experiment Setup}
\label{sec:experiment}

The Methodology section defined the four phases of \my and the design rationale for each component.
This section instantiates those design decisions as a concrete evaluation.
Specifically, it describes the $D^3$ dataset used as the benchmark, the baselines compared against, and the implementation and configuration details needed to reproduce the experiments.

\subsection{The Dockerfile Drift Dataset}
\label{sec:setup_dataset}

Using the \ex pipeline described in Section~\ref{sec:setup_extractor}, we construct the Dockerfile Drift Dataset, $D^3$, a benchmark for evaluating Dockerfile repair techniques.

We collected repositories from GitHub with at least 500 stars, created after 2020, covering nine popular programming languages: Python, C\texttt{++}, Java, C, C\#, JavaScript, SQL, Go, and Pascal.
This yielded 13,360 repositories, of which 1,775 contained Dockerfiles.
From these, \ex identified 2,616 drift instances over a three-month collection window ending in September 2025.
The GitHub Actions platform archives workflow build logs on a 90-day rotation cycle~\cite{github2025artifact}, so the dataset can be extended by repeating the collection process or broadening the repository filters.

From the 2,616 candidates, we retained only instances satisfying three reproducibility criteria: the source repository could be locally cloned, the failing commit could be checked out, and the Dockerfile referenced by the drift instance existed at that commit.
After this validation, the final $D^3$ dataset consists of 1,040 instances.
Table~\ref{tab:dataset_comparison} compares $D^3$ with existing Dockerfile datasets.
Unlike prior static-snapshot datasets, $D^3$ captures the complete build configuration required to reproduce each instance locally.
Specifically, 575 instances involve dynamic build arguments, and 869 instances specify a non-default build architecture.
Among the architecture-specified instances, 259 target two platforms, 27 target three platforms, and 18 target four platforms.

\begin{table}[t]
\centering
\caption{Statistical overview of the 1,040 $D^3$ instances across build structure,
Dockerfile complexity and co-change scope.}
\label{tab:d3_dataset_statistics}
\setlength{\tabcolsep}{5pt}
\begin{tabular}{@{}lrr@{}}
\toprule
\textbf{Metric} & \textbf{Count / Value} & \textbf{\%} \\
\midrule
\multicolumn{3}{l}{\textit{Build stages (\#\texttt{FROM})}} \\
1 stage  & 257 & 24.71 \\
2 stages & 484 & 46.54 \\
3 stages & 162 & 15.58 \\
4+ stages & 137 & 13.17 \\
\midrule
\multicolumn{3}{l}{\textit{Dockerfile structure (median / mean; min--max)}} \\
$\#$Non-comment instructions    & 27 / 37.63\ \ (6--264) & -- \\
$\#$\texttt{RUN} instructions   & 7 / 8.07\ \ (0--57)    & -- \\
$\#$\texttt{COPY}+\texttt{ADD} instructions & 5 / 7.77\ \ (0--162) & -- \\
Distinct base image references & 234 & -- \\
\midrule
\multicolumn{3}{l}{\textit{Co-change: files changed per failing commit}} \\
0 file    &   1 &  0.10 \\
1 file    & 453 & 43.56 \\
2--5 files  & 319 & 30.67 \\
5+ files &  267 &  25.66 \\
\bottomrule
\end{tabular}
\end{table}

Table~\ref{tab:d3_dataset_statistics} characterizes the structural properties of these instances across three dimensions.
\textbf{Build structure.}
75.29\% of instances involve multi-stage Dockerfiles with two or more \texttt{FROM} instructions; two-stage builds are the most common form with 484 instances.
This prevalence directly motivated the stage-aware design of \my's Context Profiler, which maintains an independent context state per build stage and resolves cross-stage ``\texttt{COPY --from}'' dependencies.
\textbf{Dockerfile complexity.}
Instances span a wide range, with a median of 27 non-comment instructions and a median of 7 \texttt{RUN} instructions per Dockerfile, confirming that $D^3$ covers both compact and complex build configurations.
\textbf{Co-change scope.}
In 43.56\% of instances, the failing commit modifies only one file, providing a clean repair signal.
In 25.66\% of instances, the failing commit changes more than 5 files.
These high-co-change cases correspond to stale drift scenarios where the most recent diff is a poor indicator of the root cause, and where the CDG's structural analysis provides the primary repair signal independently of the diff content; they are the focus of the stale-drift robustness analysis in Section~\ref{sec:results}.

\begin{table*}[t]
\centering
\caption{%
  Three-level taxonomy of the 1,040 Dockerfile drift root causes in $D^3$.
  \textbf{Bold} rows give the high-level category (L1) with its subtotal.
  Indented rows give the sub-category (L2) and specific root cause (L3) with individual
  counts.%
}
\label{tab:taxonomy_drift_causes}
\resizebox{\linewidth}{!}{
\begin{threeparttable}
\begin{tabular}{p{4cm}p{8.3cm}rr}
\toprule
\textbf{Sub-category (L2)} & \textbf{Specific root cause (L3)} & \textbf{Count} & \textbf{\%} \\
\midrule
\multicolumn{2}{l}{\textbf{Application Dependency \& Build}} & \textbf{680} & \textbf{65.4} \\
\quad Python ecosystem
  & \texttt{uv}: dependency sync \& operational errors                        & 201 & 19.3 \\
  & \texttt{pip}: installation errors (version conflicts, package not found)  & 122 & 11.7 \\
  & Python package build failures (\texttt{setup.py}, wheel)                 &  17 &  1.6 \\[3pt]
\quad JavaScript ecosystem
  & \texttt{npm}: build/install/CI failures                                   & 142 & 13.7 \\
  & \texttt{pnpm}: build failures                                             &   8 &  0.8 \\[3pt]
\quad Go ecosystem
  & Go: build, install, and dependency download failures                      &  82 &  7.9 \\[3pt]
\quad Other build systems
  & Other tools (\texttt{yarn}, poetry, Gradle, Maven, Cargo, etc.)           &  96 &  9.2 \\
  & Makefile execution failures                                               &  12 &  1.2 \\
\midrule
\multicolumn{2}{l}{\textbf{System \& Environment}} & \textbf{157} & \textbf{15.1} \\
\quad OS package management
  & Missing system dependencies (\texttt{gcc}, \texttt{ffmpeg}, \texttt{libssl}) & 77 & 7.4 \\
  & OS package manager errors (\texttt{apt-get}, \texttt{apk add})            &  35 &  3.4 \\[3pt]
\quad Runtime environment
  & Python version incompatibility                                            &  30 &  2.9 \\[3pt]
\quad External resources
  & File download errors (\texttt{wget}, \texttt{curl})                       &  15 &  1.4 \\
\midrule
\multicolumn{2}{l}{\textbf{Dockerfile-specific}} & \textbf{31} & \textbf{3.0} \\
\quad Image specification
  & Base image errors (not found, auth error, invalid tag)                    &  21 &  2.0 \\[3pt]
\quad Context specification
  & File copy/add failure (\texttt{COPY}, \texttt{ADD})                       &  10 &  1.0 \\
\midrule
\multicolumn{2}{l}{\textbf{Script \& Command}} & \textbf{30} & \textbf{2.9} \\
\quad Custom execution
  & Developer custom script failure                                           &  25 &  2.4 \\[3pt]
\quad Instruction syntax
  & \texttt{RUN} command syntax error                                         &   5 &  0.5 \\
\midrule
\multicolumn{2}{l}{\textbf{CI Infrastructure}} & \textbf{112} & \textbf{10.8} \\
\quad Build environment
  & CI cache interaction error                                                &  52 &  5.0 \\
  & Insufficient disk space during image export/build                        &  23 &  2.2 \\[3pt]
\quad Registry \& distribution
  & Docker image push/upload failure                                          &  37 &  3.6 \\
\midrule
\multicolumn{2}{l}{\textbf{Software Internal}$^\dagger$} & \textbf{30} & \textbf{2.9} \\
\quad Software-internal
  & Build failure unrelated to Dockerfile (logic bug, test failure)           &  27 &  2.6 \\[3pt]
\quad Undetermined
  & Cannot determine cause from log (general error code, no useful output)    &   3 &  0.3 \\
\midrule
\multicolumn{2}{r}{\textbf{Total}} & \textbf{1040} & \textbf{100.0} \\
\bottomrule
\end{tabular}
\begin{tablenotes}
  \small
  \item[$\dagger$] Failures in this category cannot be resolved by Dockerfile modification alone.
\end{tablenotes}
\end{threeparttable}
}
\end{table*}

Not all 1,040 instances admit a Dockerfile-only repair.
The Docker build process can itself invoke software build steps, \textit{e.g.}, \texttt{go build} within a \texttt{RUN} instruction.
Existing studies find that 26.5\%--37\% of CI builds fail due to internal software errors~\cite{8453189,10.1145/2568225.2568255}.
Such failures embedded within Dockerfile builds cannot be resolved by modifying the Dockerfile alone and represent an inherent difficulty in the benchmark.

Table~\ref{tab:taxonomy_drift_causes} reveals three patterns that define the repair challenge.
Application build-tool failures within \texttt{RUN} instructions dominate at 65.4\%.
The three largest contributors are \texttt{uv}, \texttt{npm}, and \texttt{pip}, underscoring how frequently dependency-resolution errors drive Dockerfile drift.
By contrast, only 31 instances fail at the Dockerfile instruction level via malformed \texttt{COPY}, \texttt{ADD}, or \texttt{FROM} directives, the class of error that syntax-based linters can detect.
A further 142 instances lie outside the scope of Dockerfile repair.
Specifically, 112 stem from CI infrastructure failures, such as cache interaction, disk exhaustion, registry push, and 30 from software-internal logic errors.
The remaining 898 instances are, in principle, addressable by Dockerfile modification.
For all but the 31 instruction-level cases, a correct repair requires knowing which repository files and environment variables each \texttt{RUN} instruction depends on at build time, not just what the instruction text states.

To keep the repairing process consistent with the original GitHub Action, the $D^3$ dataset records the complete build configuration for each instance, including the exact commit hash, Dockerfile path, build context path, dynamic build arguments, and target platforms.
These parameters are identical to those used in the original CI environment, ensuring that a local build initiated with the provided configuration is equivalent to the GitHub Actions run.
The dataset, the \ex tool, and all analysis scripts will be made publicly available upon publication.

\subsection{Experimental Environment}

All experiments ran on a server equipped with dual AMD EPYC 7543 32-core processors (128 threads total), 1.0\,TB of RAM, 21\,TB of disk storage, and Ubuntu 22.04.4 LTS.

The experiments with LLM are conducted using DeepSeek-V3~\cite{liu2024deepseek}. We built a Docker build infrastructure with 16 isolated \texttt{docker buildx} builders using the \texttt{docker-container} driver.
This driver provides full build isolation and supports multi-platform image construction via \texttt{buildx}.
The 16 builders ran in parallel to reduce the total wall-clock time of the experiment.

For each drift instance, we checked out the corresponding commit to reconstruct the exact source state.
A repair was counted as successful if the patched Dockerfile built without error using the precise build arguments and platform specifications from $D^3$.
Each build attempt had a 30-minute timeout, consistent with prior work~\cite{9402688,11029932}.
We pruned each builder's build cache every 8 builds to prevent cross-contamination between experiments.

\section{Experimental Results}
\label{sec:results}

Section~\ref{sec:experiment} defined the $D^3$ benchmark, and this section reports what the experiments reveal across three research questions, each targeting a distinct dimension of \my's behavior.

\begin{itemize}
    \item \textbf{RQ1: How effective is \my in repairing real-world Dockerfile drift?}
    This question measures the absolute and relative repair capability of \my, verifying whether context-aware dependency modeling translates into higher repair rates on a realistic benchmark.

    \item \textbf{RQ2: How efficient and robust is \my in repairing Dockerfile drift?}
    This question examines two practical characteristics.
    The first is efficiency: whether \my consumes fewer tokens than competing LLM-based methods.
    The second is robustness: whether its performance degrades more slowly as drift instances become ``stale'' over successive commits.
    These two properties determine whether \my remains useful in realistic CI/CD deployments where build failures may persist across multiple commits.

    \item \textbf{RQ3: How do individual components of \my contribute to its overall effectiveness?}
    This question uses controlled ablation to attribute performance improvements to components of \my, verifying that each design choice is independently justified.
\end{itemize}

\subsection{RQ1: Overall Repair Effectiveness}
\label{sec:results_rq1}

\begin{table}[b]
    \caption{Per-method outcome breakdown on the 653-instance $D^3$ benchmark.
    $|\mathcal{S}|$: builds successfully after patching; $|\mathcal{F}_r|$: no valid patch generated; $|\mathcal{F}_b|$: patch generated but build fails; $|\mathcal{T}|$: build timeout; $\text{RR}$: repair rate.}
    \label{tab:ori_results}
    \begin{threeparttable}
    \begin{tabular}{lrrrrrr}
        \toprule
        \textbf{Method} & $N$ & $|\mathcal{S}|$ & $|\mathcal{F}_r|$ & $|\mathcal{F}_b|$ &
          $|\mathcal{T}|$ & $\text{RR}$ (\%) \\
        \midrule
        Parfum              & 653 &  61 & 35 & 546 & 11 &  9.34 \\
        FlakiDock           & 653 & 133 & 48 & 462 & 10 & 20.37 \\
        FlakiDock-DS-V3\tnote{\dag}  & 653 &   0 & 41 & 612 &  0 &  0.00 \\
        FlakiDock-DS-V3\tnote{\ddag} & 653 & 185 & 43 & 397 & 28 & 28.33 \\
        Vanilla-LLM         & 653 & 186 & 58 & 380 & 29 & 28.48 \\
        \textbf{\my}        & \textbf{653} & \textbf{230} & \textbf{0} & \textbf{399} &
          \textbf{24} & \textbf{35.22} \\
        \bottomrule
    \end{tabular}
    \begin{tablenotes}
        \item[\dag]  FlakiDock with DeepSeek-V3, original output handling from FlakiDock.
        \item[\ddag] FlakiDock with DeepSeek-V3 and \my's output-parsing.
    \end{tablenotes}
    \end{threeparttable}
\end{table}

\begin{table*}[t]
\centering
\caption{Per-category repair rates on the 653-instance $D^3$ benchmark.
  Rates are computed at the L1 taxonomy level as defined in Table~\ref{tab:taxonomy_drift_causes}.
  }
\label{tab:per_category_rr}
\begin{tabular}{lrrrrr}
\toprule
\textbf{Category (L1)} & $N$ &
  \textbf{Parfum} & \textbf{FlakiDock-DS-V3} & \textbf{Vanilla-LLM} & \textbf{\my} \\
\midrule
Application Dependency \& Build & 346 &  4.9\% & 19.1\% & 20.2\% & \textbf{25.1\%} \\
System \& Environment           & 135 & 14.1\% & 51.9\% & 51.9\% & \textbf{60.0\%} \\
Dockerfile-specific             &  30 &  0.0\% & 50.0\% & 53.3\% & \textbf{60.0\%} \\
Script \& Command               &  29 &  0.0\% &  3.4\% & 17.2\% & \textbf{27.6\%} \\
CI Infrastructure     &  85 & 28.2\% & 36.5\% & 28.2\% & \textbf{40.0\%} \\
Software Internal     &  28 &  3.6\% &  \textbf{7.1\%} &  3.6\% &   \textbf{7.1\%}          \\
\midrule
\textbf{All}                    & 653 &  9.3\% & 28.3\% & 28.5\% & \textbf{35.2\%} \\
\bottomrule
\end{tabular}
\end{table*}

\begin{figure*}[b]
  \includegraphics[width=\linewidth]{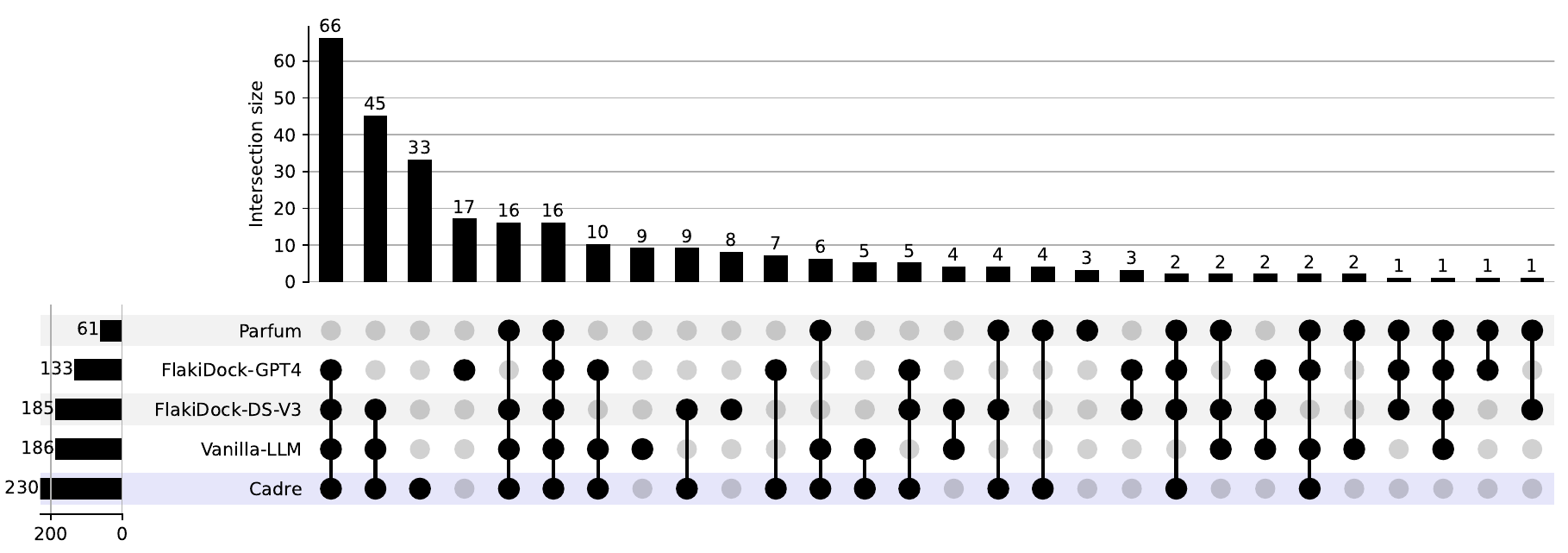}
  \caption{Upset plot of successfully repaired cases per method.
  Horizontal bars show individual method totals.
  Intersection bars show the number of cases repaired by exactly the indicated combination of methods.
  The third intersection bar represents cases repaired by \my and by no other method.}
  \label{fig:upset}
\end{figure*}

To examine the repair capability of each method, we evaluated all tools on the 653-instance executable benchmark derived from $D^3$.
The 653-instance subset excludes the 387 instances that consistently timed out across all tools due to network-bound factors that are unrelated to the logical correctness of a repair. These cases typically involve multi-gigabyte base image downloads like CUDA images or large dependency installations like PyTorch with cuda-runtime.
This filtering is applied uniformly to all methods, and the resulting benchmark is still larger than the 344-instance dataset used in the FlakiDock evaluation~\cite{11029932}.

Table~\ref{tab:ori_results} reports the outcome breakdown for each method.
We classify each repair attempt into one of four disjoint categories:
$\mathcal{S}$ means successful repair, that the patched Dockerfile builds without error.
$\mathcal{F}_r$ means a repair failure, that no valid Dockerfile is generated.
$\mathcal{F}_b$ means a build failure, that a patch is produced, but the build fails.
$\mathcal{T}$ means the build process exceeds the time limit we defined.
For $N$ instances, the Repair Rate is $\text{RR} = |\mathcal{S}| / N$.

\my achieves $\text{RR} = 35.22\%$, successfully repairing 230 of the 653 benchmark instances.
\my outperforms the best LLM-based baseline, Vanilla-LLM, by 6.74 percentage points, and the rule-based \texttt{Parfum} by 25.88 percentage points.
Expressed as a ratio, \my's repair rate is $3.77\times$ that of \texttt{Parfum} and $1.24\times$ that of the best LLM-based baseline.

The comparison against \texttt{FlakiDock} warrants careful interpretation.
When \texttt{FlakiDock} is run with DeepSeek-V3 under its original output-handling procedure, $|\mathcal{S}| = 0$ because excessive context length causes the LLM to deviate from the expected output format.
After applying \my's output-parsing procedure to enable fair comparison, \texttt{FlakiDock-DS-V3} reaches $\text{RR} = 28.33\%$.
The performance ceiling of the knowledge-base retrieval strategy is apparent: the pre-constructed example pairs do not cover the diversity of build contexts in $D^3$, whereas \my extracts context directly from the live repository without any prior knowledge.

A notable characteristic of \my is $|\mathcal{F}_r| = 0$: the framework always produces a candidate patch, regardless of whether the patch succeeds.
Vanilla-LLM and the FlakiDock variants each produce $|\mathcal{F}_r|$ between 41 and 58, indicating that their prompts periodically trigger refusals or format failures due to the model's token limits.

Figure~\ref{fig:upset} visualizes the intersection of successfully repaired instances across all methods.
\my repairs the largest set of unique instances that no other method repairs, confirming that its context-aware approach recovers cases that alternative strategies cannot address.

\begin{table*}[t]
\centering
\caption{Repair rate distribution by Dockerfile build-stage count.}
\label{tab:repair_by_stage}
\begin{tabular}{lrrrrr}
\toprule
\textbf{Method} & \textbf{1} ($N$=200) & \textbf{2} ($N$=171) & \textbf{3} ($N$=155) &
  \textbf{4+} ($N$=127) & \textbf{All} ($N$=653) \\
\midrule
Parfum          & 24/200 (12.0\%) & 25/171 (14.6\%) &  2/155 (1.3\%)  & 10/127 (7.9\%)  & 61/653 (9.3\%)   \\
FlakiDock       & 20/200 (10.0\%) & 82/171 (48.0\%) & 21/155 (13.6\%) & 10/127 (7.9\%)  & 133/653 (20.4\%) \\
FlakiDock-DS-V3 & 30/200 (15.0\%) & 103/171 (60.2\%)& 35/155 (22.6\%) & 17/127 (13.4\%) & 185/653 (28.3\%) \\
Vanilla-LLM     & 33/200 (16.5\%) & 111/171 (64.9\%)& 29/155 (18.7\%) & 13/127 (10.2\%) & 186/653 (28.5\%) \\
\textbf{\my}    & \textbf{45/200 (22.5\%)} & \textbf{121/171 (70.8\%)} &
  \textbf{42/155 (27.1\%)} & \textbf{22/127 (17.3\%)} & \textbf{230/653 (35.2\%)} \\
\bottomrule
\end{tabular}%
\end{table*}

\begin{figure*}[b]
  \centering
  \includegraphics[width=0.9\linewidth]{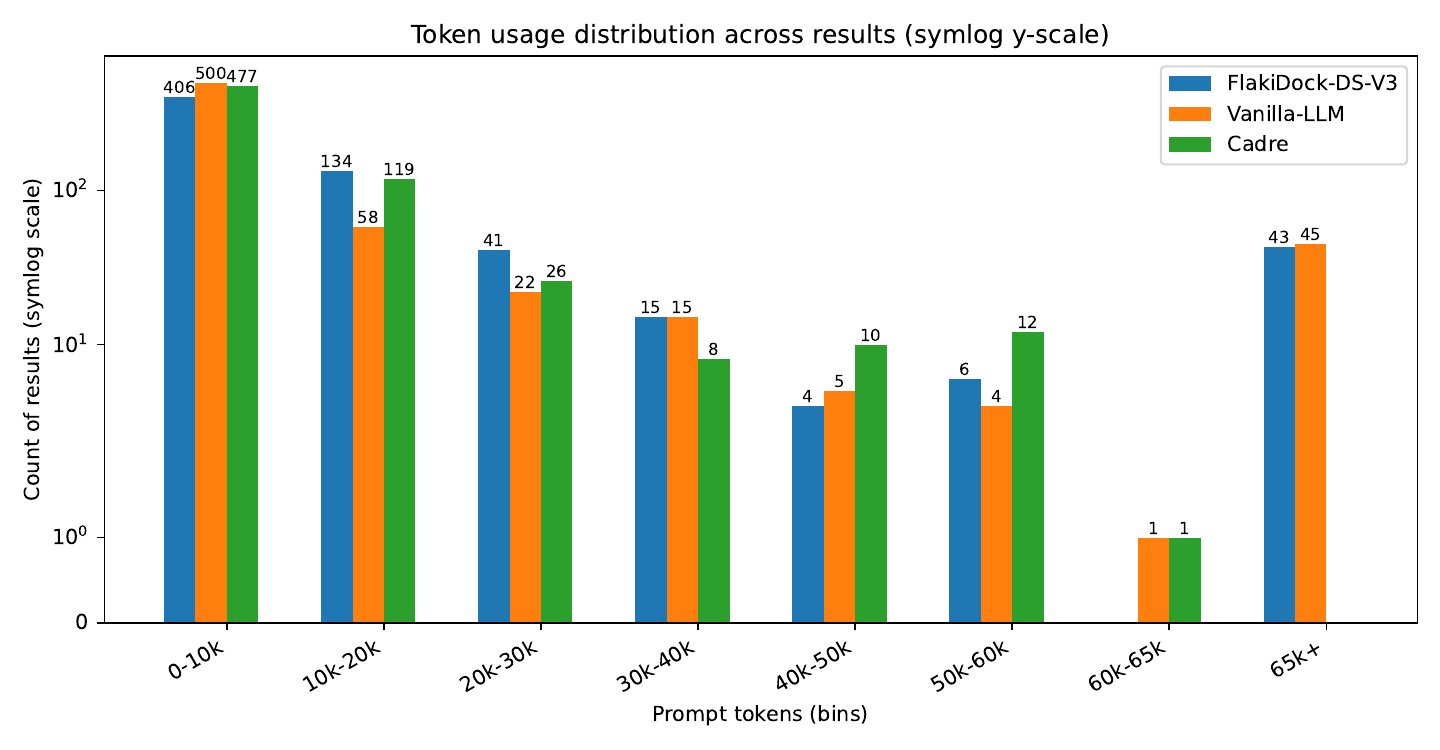}
  \caption{Distribution of prompt token counts per repair attempt.
  The y-axis uses a symmetric log scale.
  Counts above 65,535 tokens exceed the DeepSeek-V3 context limit~\cite{deepseek-v3}.}
  \label{fig:token_usage}
\end{figure*}

\textbf{Per-category repair analysis.}
We mapped each benchmark instance to the taxonomy in Table~\ref{tab:taxonomy_drift_causes} and computed $\text{RR}$ per L1 category.
Table~\ref{tab:per_category_rr} presents the results.
\my achieves the highest rate in every fixable category, with System \& Environment and Dockerfile-specific both reaching 60.0\%.
Within System \& Environment, missing-dependency failures reach 78.1\% under \my, the highest rate of any sub-category, because the CDG directly infers which system packages each \texttt{RUN} instruction requires through build-system-aware parsers.
While rule-based and retrieval-based methods stall near zero on Script \& Command failures, \my reaches 27.6\% by tracing script invocations through CDG file-dependency edges.
In the Application Dependency category, \my achieves 17.0\% on less common build tools such as Gradle, Maven, and Cargo, four times Vanilla-LLM's 4.3\%, because live CDG extraction covers tools absent from FlakiDock's training corpus.
The one exception is OS package manager errors, where FlakiDock's 52.9\% exceeds \my's 26.5\%, reflecting the template coverage that retrieval-based repair provides for \texttt{apt-get} and \texttt{apk add} patterns.
The elevated CI Infrastructure rates are a measurement artifact: registry push failures occur after the Docker build succeeds, so any produced patch trivially passes local verification.

\textbf{Multi-stage Dockerfile analysis.}
Table~\ref{tab:repair_by_stage} disaggregates $\text{RR}$ by the number of \texttt{FROM} instructions, which determines the number of independent build stages and the depth of cross-stage dependency tracking required.

On two-stage builds, \my achieves 70.8\% and Vanilla-LLM reaches 64.9\%, the highest repair rates for any stage count.
The canonical build-plus-runtime structure, common to Go, Node.js, and Python projects, gives LLMs strong prior familiarity with repair patterns for this configuration.
A more telling pattern emerges at higher stage counts.
\my's relative advantage over Vanilla-LLM grows from 36\% at one stage to 45\% at three stages and to 70\% at four or more.
This widening confirms that the CDG's stage-aware tracking provides increasing value as cross-stage dependencies accumulate with structural complexity.

\begin{figure*}[b]
  \includegraphics[width=\linewidth]{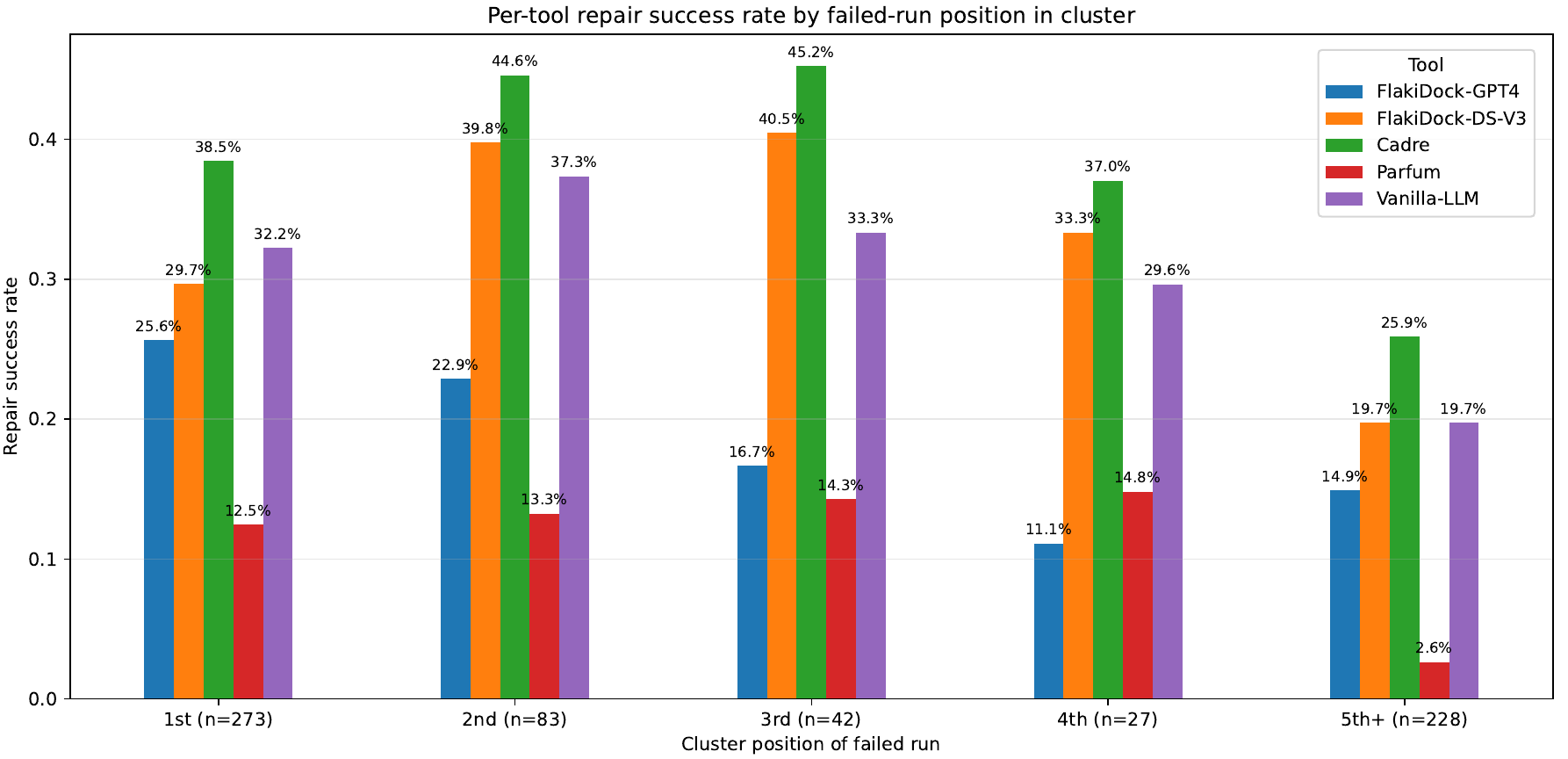}
  \caption{Repair success rate by cluster failure distance.
  Higher distance means the drift has persisted through more subsequent commits, making the most recent diff increasingly misleading.}
  \label{fig:cluster}
\end{figure*}

\textbf{Failure analysis:} The 423 unrepaired \my attempts ($|\mathcal{F}_b| + |\mathcal{T}|$) distribute across identifiable failure families rather than forming a uniform residual.
To characterize these families, we performed a log-signature clustering analysis.
For each case, we extracted the most diagnostically specific line from the post-patch build log, preferring inner build-system error messages over the generic Docker wrapper line when both were present.
We then normalized instance-specific tokens such as paths, hashes, URLs, and numerals, grouped identical normalized signatures into exact buckets, and iteratively merged buckets whose representative signatures exceeded a Jaccard--sequence similarity of 0.88.

The analysis produced 75 clusters, with a strongly head-concentrated distribution: 32 clusters (42.7\%) are singletons, yet the five largest clusters account for 231 cases, which are 54.6\% of all unrepaired attempts.
The dominant cluster, comprising 57 cases, groups Maven and Java package construction failures that the patch did not resolve.
The second-largest pattern, at 52 cases, consists of opaque \texttt{RUN}-command exits in which the post-patch log records only that a package-manager or build-tool command exited non-zero, without a more specific inner diagnostic.
Affected commands span package managers such as \texttt{npm}, \texttt{pnpm}, and \texttt{uv}, as well as build tools such as \texttt{go generate}.
Python dependency resolution accounts for two further clusters totalling 85 cases (20.1\%), in which the representative signature in both is a pip version-constraint unsatisfiability error.
The remaining notable clusters cover 37 cases of missing build-context files, where \texttt{BuildKit} reports a checksum failure for a file absent from the build context, \textit{e.g.}, \texttt{uv.lock},; 24 hard timeouts; and 10 failures caused by base images unavailable from a local registry.

Taken together, the clustering results show that \my's unresolved cases are concentrated in three structurally distinct failure families: build-tool constraint failures that require runtime package-registry state unavailable at static analysis time; context-file mismatches where the patch identifies but does not fully resolve the dependency on a newly added file; and environment-dependent replay failures
caused by infrastructure limits rather than Dockerfile logic.
None of these families arises from prompt-format failures or context overflow, the failure modes that the two-step workflow is designed to prevent.

\begin{takeaway}
\noindent\textbf{Answer to RQ1.}
\my achieves 35.22\% on $D^3$, outperforming Vanilla-LLM by 6.74 percentage points
and \texttt{Parfum} by 25.88 percentage points.
The largest per-category gains appear in Script \& Command failures with 27.6\% compared to \ 0.0\% for \texttt{Parfum}, and System \& Environment failures with 60.0\%, with the relative advantage over Vanilla-LLM growing to 70\% at four or more build stages.
The unresolved cases concentrate on build-tool constraint failures, context-file mismatches, and environment-dependent replay failures that fall outside the reach of static dependency analysis.
\end{takeaway}

\subsection{RQ2: Efficiency and Robustness}

Beyond repair accuracy, RQ2 examines two practical characteristics: how efficiently \my uses the LLM's token budget, and how well it handles drift instances that have persisted across commits.

\subsubsection{Token Efficiency}

An important property of any LLM-based tool is the size of the prompts it constructs.
Oversized prompts increase operational cost, consume a larger fraction of the model's context window, and risk diluting the model's attention across irrelevant content.
We measured the prompt token count for each repair attempt by every LLM-based method.

Figure~\ref{fig:token_usage} shows the token distribution.
\my holds the prompt below 30k tokens in 95.25\% of cases.
In contrast, Vanilla-LLM and FlakiDock-DS-V3 produce 43 and 45 instances that exceed DeepSeek-V3's 65k-token context limit.
These over-limit cases account for the non-zero $|\mathcal{F}_r|$ counts in Table~\ref{tab:ori_results}: when the prompt exceeds the context window, the LLM can't generate a valid patch.

The efficiency advantage of \my is a direct consequence of the Context Inquiry step, which serves as a filter before any file contents are loaded.
By first selecting the most relevant files from the CDG, \my avoids supplying the LLM with large, irrelevant source files, producing shorter and more focused prompts than approaches that supply all available information upfront.

\subsubsection{Robustness to Stale Drift}

A unique feature of $D^3$ is the presence of stale drift instances: cases where a build failure introduced at one commit persists across several subsequent commits.
In these later builds, the code-change diffs are unrelated to the root cause of the failure.
Tools that rely exclusively on the most recent diff for repair context are misled by this noise.
In contrast, a tool that analyzes the structural dependency graph can trace the failure back to its origin regardless of how many commits have elapsed.

To measure this, we define the \textbf{Cluster Failure Distance} as the number of failed builds between the commit that introduced the drift and the build being repaired.
We identify clusters of related failures by grouping build logs using character 5-gram Jaccard similarity with a threshold of $\tau = 0.8$.
Within each cluster, the position of each failure, \textit{i.e.}, 1st, 2nd, 3rd, \textit{etc.}, gives its distance.

Figure~\ref{fig:cluster} shows $\text{RR}$ as a function of cluster failure distance.
All methods show a declining trend as distance increases, confirming that stale drifts are harder to repair.
\my maintains a higher success rate than all baselines at every distance level.
At the first position, \textit{i.e.}, the fresh failures, \my achieves 38.5\%, compared to 32.2\% for Vanilla-LLM.
At the fifth position and beyond, \textit{i.e.}, stale failures, \my achieves 25.9\%, compared to 19.7\% for Vanilla-LLM.
This gap is proportionally larger than at the first position, indicating that the CDG's structural analysis provides a repair signal that remains informative when the recent diff is noisy.
Diff-only approaches, by contrast, degrade more steeply as commit distance grows.

\begin{takeaway}
\textbf{Answer to RQ2.}
The Context Inquiry step keeps every prompt within the model's operational range, producing $|\mathcal{F}_r| = 0$ while LLM-based baselines fail to generate a valid patch in 41--58 cases due to context overflow.
On stale drift persisting five or more commits, \my achieves 25.9\% against Vanilla-LLM's 19.7\%.
The proportionally larger gap than at fresh positions confirms that the CDG's structural signal degrades more slowly than diff-based context as failures age.
\end{takeaway}

\subsection{RQ3: Ablation on Individual Components}
\label{sec:results_rq3}

RQ3 uses controlled ablation to verify that the CDG and the two-step repair workflow independently contribute to \my's performance.
The two ablation variants form a gradient from no CDG to the full framework.
Table~\ref{tab:ablation} presents the results.

\textbf{Removing the CDG (\my-CDG).}
This variant removes the CDG.
The LLM receives only the code-change list and the failure log, processed through the Context-refined Repair workflow.
This variant tests whether any graph-structured context, as opposed to raw diffs and logs, contributes to repair performance.

Without any graph-structured context, $\text{RR}$ drops from 35.22\% to 30.93\%.
At this level, \my degenerates to a workflow-augmented version of Vanilla-LLM: it still uses the two-step repair process but lacks the structural context that guides the first step.
The 4.29-percentage-point gap confirms that the CDG contributes independently of the repair workflow.

\begin{table}[b]
    \centering
    \caption{Ablation study on the 653-instance $D^3$ benchmark.
    Each variant removes one structural layer of the full \my framework.
    Column definitions follow Table~\ref{tab:ori_results}.}
    \label{tab:ablation}
    \begin{threeparttable}
    \begin{tabular}{lrrrrrr}
        \toprule
        \textbf{Method} & $N$ & $|\mathcal{S}|$ & $|\mathcal{F}_r|$ & $|\mathcal{F}_b|$ &
          $|\mathcal{T}|$ & $\text{RR}$ (\%) \\
        \midrule
        \my-CDG\tnote{\dag}  & 653 & 202 &  2 & 413 & 36 & 30.93 \\
        \my-CRR\tnote{\ddag} & 653 & 206 & 44 & 373 & 36 & 31.55 \\
        Vanilla-LLM          & 653 & 186 & 58 & 380 & 29 & 28.48 \\
        \textbf{\my}         & \textbf{653} & \textbf{230} & \textbf{0} & \textbf{399} &
          \textbf{24} & \textbf{35.22} \\
        \bottomrule
    \end{tabular}
    \begin{tablenotes}
        \item[\dag]  Removes the CDG; supplies only code-change diffs and error log.
        \item[\ddag] Retains the CDG; removes the two-step repair workflow (single-shot).
    \end{tablenotes}
    \end{threeparttable}
\end{table}

\textbf{Removing the two-step workflow (\my-CRR).}
This variant retains the complete CDG but removes the two-step agentic workflow.
It operates in a single shot, supplying the LLM with all available information at once: the error log, the full CDG, the original Dockerfile, and the raw diffs of all changed files.
This variant tests whether the iterative context-selection step is necessary or whether a richer single-shot prompt achieves equivalent performance.

Without the iterative context-selection step, $\text{RR}$ drops to 31.55\% and $|\mathcal{F}_r|$ rises from 0 to 44.
The increase in $|\mathcal{F}_r|$ is the direct consequence of context overload.
When all available information is supplied in a single prompt, a substantial fraction of prompts exceed the 65k-token context limit.
This confirms that the Context Inquiry step is not merely a performance optimization but a necessary mechanism for keeping prompts within the model's operational range.

\begin{figure}[t]
  \centering
  \includegraphics[width=\linewidth]{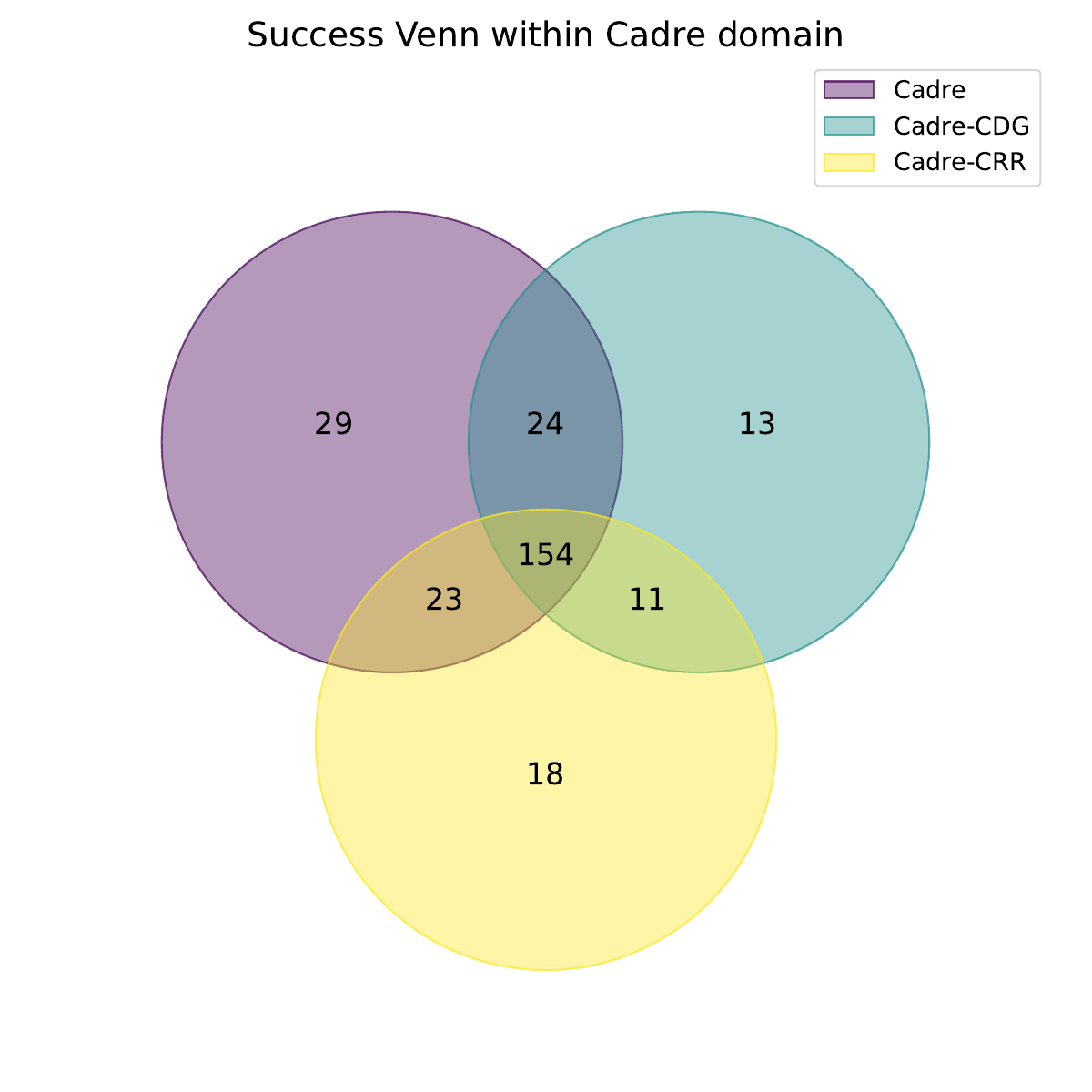}
  \caption{Venn diagram of successfully repaired cases across \my ablation variants.
  The region unique to the full \my represents cases that require all components to be simultaneously active.}
  \label{fig:ablation_venn}
\end{figure}

\textbf{Component interaction.}
The CDG and the two-step workflow are super-additive in their combined contribution.
Relative to Vanilla-LLM, the CDG alone (\my-CRR) contributes 3.07 percentage points and the workflow alone (\my-CDG) contributes 2.45 percentage points.
Under an independence assumption, their combined gain would be 5.52 percentage points.
The observed gain is 6.74 percentage points, exceeding the additive expectation by 1.22 percentage points.
This surplus indicates mutual reinforcement.
The CDG provides higher-quality input to the Context Inquiry step.
In turn, the Context Inquiry step allows the CDG's structural information to be applied without exhausting the model's token budget.

Figure~\ref{fig:ablation_venn} confirms that the components are not redundant.
\my-CDG and \my-CRR each repair a distinct subset of cases that the other variant misses.
The region exclusive to the full \my system represents failures that require both structural context and selective loading to resolve, cases that neither variant can address alone.

\begin{takeaway}
\textbf{Answer to RQ3.}
Both the CDG and the two-step workflow contribute independently, with CDG removal reducing $\text{RR}$ by 4.29 percentage points and workflow removal by 3.67 points while raising $|\mathcal{F}_r|$ from 0 to 44.
The combined gain over Vanilla-LLM exceeds the additive expectation by 1.22 percentage points, indicating that the two components are mutually reinforcing.
The CDG provides higher-quality evidence for the Context Inquiry step, which in turn keeps the CDG's structural information within the model's context budget.
\end{takeaway}

\begin{figure*}[t]
\centering

\begin{tcolorbox}[
  enhanced,
  colback=white,
  colframe=gray!50!black,
  boxrule=1.2pt,
  arc=4pt,
  left=3pt,
  right=3pt,
  top=2pt,          
  bottom=2pt,
  title={\textbf{Case Study: PeerBanHelper}},
  coltitle=white,
  colbacktitle=gray!50!black,
  fonttitle=\sffamily\bfseries\normalsize,
  attach boxed title to top left={yshift=-4pt, xshift=6pt},
  boxed title style={
    arc=2pt,
    boxrule=0pt,
    left=5pt,
    right=5pt,
    top=0pt,
    bottom=0pt,
    colback=gray!50!black,
  },
]

\begin{methodlisting}{Original (failing)}{bgorig}{origgray}
RUN apk add git && \
    mvn -B clean package --file pom.xml -T 1.5C -P thin-sqlite-packaging
\end{methodlisting}

\begin{methodlisting}{Vanilla-LLM (incorrect)}{bgvanilla}{vanillared}
RUN apk add git && \
    mvn -B clean package --file pom.xml -T 1.5C -P thin-sqlite-packaging \
(*\plusvanilla*)        -Dmaven.wagon.http.ssl.insecure=true \
(*\plusvanilla*)        -Dmaven.wagon.http.ssl.allowall=true
\end{methodlisting}

\begin{methodlisting}{Cadre (correct)}{bgcadre}{cadregreen}
RUN apk add git && \
(*\pluscadre*)    mvn install:install-file                              \
(*\pluscadre*)        -Dfile=lib/sqlite-jdbc-loongarch64-3.47.0.0.jar   \
(*\pluscadre*)        -DgroupId=com.ghostchu.peerbanhelper.external-libs\
(*\pluscadre*)        -DartifactId=sqlite-jdbc-loongarch64              \
(*\pluscadre*)        -Dversion=3.47.0.0 -Dpackaging=jar && \
    mvn -B clean package --file pom.xml -T 1.5C -P thin-sqlite-packaging
\end{methodlisting}

\end{tcolorbox}

\caption{A case study from PeerBanHelper~\cite{PeerBanHelper}. The failing \texttt{RUN} instruction and the two repairs.
\textbf{\textcolor{cadregreen}{Cadre}} correctly installs the local JAR before building.
\textbf{\textcolor{vanillared}{Vanilla-LLM}} misdiagnoses the failure as an SSL connectivity issue.}
\label{lst:case_repair}
\end{figure*}

\subsection{Case Study: Context-Dependent Drift Repair}
\label{sec:case_study}

We trace a representative drift instance from PeerBanHelper~\cite{PeerBanHelper}, an open-source BitTorrent peer management application, to illustrate how \my's file-level dependency tracking enables repairs that context-free approaches cannot produce.

\textbf{The drift.}
The project introduced a LoongArch64-specific SQLite JDBC driver as a local Maven dependency by adding ``\texttt{lib/sqlite-jdbc-loongarch64-3.47.0.0.jar}'' to the repository.
Because the artifact is not available on Maven Central, it must be installed into the local Maven repository before the build.
The Dockerfile was not updated accordingly, causing the Maven build stage to fail.

\textbf{How \my traces the repair evidence.}
The ``\texttt{COPY . /build}'' instruction copies the entire repository into the Maven build stage, including ``\texttt{lib/sqlite-jdbc-loongarch64- 3.47.0.0.jar}''.
\my's Context Profiler captures this file in the stage's context state $\Phi$, and the CDG records a dataflow edge from the file vertex to the ``\texttt{RUN mvn package}'' instruction vertex that follows.
During Context Inquiry, the LLM receives this CDG structure alongside the commit diff.
From the failure log and the CDG edge, the LLM identifies that the JAR is a prerequisite of the failing instruction.
The repair is then direct: insert ``\texttt{mvn install:install-file}'' to register the JAR in the local repository before the package step.

\textbf{Why context-free approaches fail.}
Without the CDG, Vanilla-LLM receives only the error log and the diff.
Maven's dependency resolution error, ``\textit{Could not find artifact \texttt{com.ghostchu.peerbanhelper.external-libs:sqlite-jdbc -loongarch64:jar:3.47.0.0}}'', is superficially indistinguishable from SSL-related network failures that arise when Maven cannot reach a remote repository.
Vanilla-LLM applies a textbook fix for SSL failures, adding certificate bypass flags that are entirely irrelevant here.
The JAR is present in the repository and is visible in the container filesystem via ``\texttt{COPY . /build}'', but this evidence is accessible only through file-level context modeling.
All other baselines fail for the same reason: no mechanism connects the newly introduced binary file to the \texttt{RUN} instruction that depends on it.

This instance represents a class of drifts where the repair evidence is a binary artifact that appears in neither the Dockerfile text nor the error log.
The CDG's dataflow edges, seeded by context-ingesting instructions such as ``\texttt{COPY . /build}'' and propagated to dependent \texttt{RUN} instructions, are the mechanism that surfaces this evidence and makes the repair tractable.

\section{Discussion}
\label{sec:discussion}

\subsection{Implications and Limitations}

The consistent pattern across all three research questions is that context structure determines repair quality more than context volume.
In RQ1, \my outperforms Vanilla-LLM despite both tools receiving the same categories of information.
In RQ2, the performance gap widens precisely where diff-based context is least reliable: at high commit distances, the CDG's structural signal remains stable while diff-only approaches degrade.
In RQ3, removing either the CDG or the two-step workflow costs more than their individual contributions would predict independently, confirming that the two components are mutually reinforcing rather than separable.
The practical consequence for tool builders is that dependency graph construction and context selection should be co-designed.
Prompt engineering addresses how information is presented; a dependency graph addresses what information is relevant, and the latter problem does not disappear as LLMs grow stronger.

Three limitations bound these findings.
First, the CDG is built from static analysis and cannot represent runtime factors such as package registry state at resolution time or network-conditional build behavior; failures caused by these dynamic factors account for a portion of the unresolved $|\mathcal{F}_b|$ cases.
Second, $D^3$ covers only public GitHub repositories, so enterprise environments with private registries, internal package mirrors, or confidential build secrets introduce failure modes not represented in the benchmark.
Third, whether the repair rates observed on Dockerfiles generalize to other IaC formats depends on the quality of static analysis achievable for those instruction semantics, which warrants a dedicated empirical study.

\subsection{Threats to Validity}
\label{sec:threats}

\textbf{Internal Validity.}
Transient environmental factors, \textit{i.e.}, network instability, resource contention, or registry unavailability, could cause a build to fail for reasons unrelated to the Dockerfile repair.
We address this through 16 isolated docker buildx builders, an automatic retry mechanism for network errors up to three attempts, and manual review of a sample of recorded failures.
The residual risk is that some failures reflect upstream registry changes between the original CI run and our local reproduction; this risk affects all compared methods equally and does not bias the relative ranking.

\textbf{External Validity.}
The dataset is drawn exclusively from public GitHub repositories across nine programming languages, so enterprise environments with private base images or organizational build secrets introduce failure modes not captured.
Among the nine languages, compiled toolchains with complex build configurations may present different repair challenges from the interpreted-language cases that dominate $D^3$.
Despite these constraints, $D^3$ is the largest real-world Dockerfile drift benchmark to date and covers a broader range of build configurations, including dynamic build arguments and multi-platform targets, than any prior dataset.

\section{Related Work}
\label{sec:related_work}

\subsection{Dockerfile Quality Analysis}

Large-scale empirical studies have characterized the Docker ecosystem at scale~\cite{cito2017empirical,xu2018mining}, tracked how Dockerfile instructions evolve as projects develop~\cite{wu2020changes,henkel2020learning}, and cataloged the maintenance challenges developers face~\cite{haque2020challenges}.
Image-level analyses further show that technical lag in base layers~\cite{zerouali2021technical} and divergence among images nominally for the same system~\cite{ibrahim2020too} sustain quality problems that instruction-level rules cannot detect.
Static analysis tools detect Dockerfile issues by matching instructions against fixed rule sets.
\texttt{Hadolint}~\cite{hadolint} flags best-practice violations, \texttt{Dockle}~\cite{dockle} targets security anti-patterns, and Docker's built-in check subsystem~\cite{docker2025build-checks} provides similar coverage at build time.
Empirical studies have cataloged Dockerfile smells, measured their prevalence, and linked them to build
failures~\cite{rosa2024fixing,ksontini2024drminer,wu2025towards}.
Both lines of work operate on the Dockerfile text alone and cannot model how instruction correctness depends on the external repository state.
A \texttt{COPY} instruction referencing a renamed source path passes all rule checks yet causes a build failure at the next CI run.
\my's Instruction-level Context Profiler addresses this gap by tracking file-level and environment-variable dependencies per instruction.

\subsection{Automated Repair of Dockerfiles}

The automated program repair literature has progressed from search-based mutation~\cite{legoues2012genprog,monperrus2018repair} to zero-shot and conversational LLM patching~\cite{xia2022alpharepair,xia2024chatrepair,bouzenia2025repairagent}.
Dockerfile-specific repair has followed the same arc from rule-based templates toward LLM-driven generation.
DockerizeMe~\cite{horton2019dockerizeme} showed that graph-based reasoning over package indices can automatically infer environment dependencies for code snippets, establishing that structured dependency representations
benefit automated environment problem-solving.
Rule-based tools such as \texttt{Parfum}~\cite{parfum2024empirical} apply template-based fixes to AST-detected smell patterns.
They achieve reliable coverage within their rule library but cannot address repository-specific failures such as renamed dependency files or changed build-tool invocations; on $D^3$, \texttt{Parfum} achieves 9.34\%.
LLM-based approaches~\cite{11025789,10336292} generate patches from error logs without explicit context modeling.
Zhu et al.~\cite{10.1145/3728870} extend instruction-level analysis to build-time efficiency rather than fault correction, demonstrating that build-process analysis is productive beyond correctness repair.
\texttt{FlakiDock}~\cite{11029932} augments LLM repair with a retrieved knowledge base of historical Dockerfile repair pairs but does not incorporate live build context, limiting coverage to patterns present in the training corpus.
None of these approaches tracks which repository files each instruction depends on or how these dependencies propagate across instruction boundaries.
\my addresses this gap through the CDG, which provides the LLM with a structured dependency representation rather than a flat list of changed files or retrieved historical examples.

\subsection{Dockerfile and IaC Benchmarks}

Prior Dockerfile datasets consist of static GitHub snapshots that capture only Dockerfile text, without the failing commit, the triggering code-change diff, or the dynamic build arguments passed to the CI runner~\cite{9402688,11025789,wu2020buildfailures}.
Cassano et al.~\cite{11029932} extracted Dockerfile-error pairs from GitHub Actions logs but omitted build arguments and platform specifications, causing 38\% of instances to fail local reproduction.
$D^3$ records the complete build configuration from actual CI runner logs, including dynamic build arguments and multi-platform specifications.
Table~\ref{tab:dataset_comparison} compares $D^3$ with prior datasets. The key differentiator is the capture of dynamic build parameters and multi-platform specifications, which are absent from all prior datasets.

\subsection{Configuration Drift in Software Evolution}

Dockerfile drift is one instance of a broader class of problems in which configuration files fall out of synchronization with the source code they configure.
Zhang et al.~\cite{zhang2021evolutionary} and Jiang et al.~\cite{e2022infrastructure} documented recurring IaC desynchronization patterns and identified the change-induced drift in cloud deployment environments. 
At the build-system level, Mukherjee et al.~\cite{mukherjee2021fixing}, Peng et al.~\cite{peng2024less}, and Sun et al.~\cite{sun2023revisiting} respectively linked dependency specification drift, build configuration complexity, and CI pipeline smells to elevated build failure rates during software evolution.
These studies characterize how drift occurs but do not address automated repair.
\my contributes an automated repair approach designed for the dependency structure of Dockerfile builds, with $D^3$ as a reproducible benchmark for measuring effectiveness under realistic evolutionary conditions.

\section{Conclusion}
\label{sec:conclusion}
 
Dockerfile drift arises when a project's Dockerfile falls out of synchronization with its evolving source code, causing build failures that existing tools cannot repair because they analyze the Dockerfile in isolation from the build context.
This paper presents \my, which addresses this gap by explicitly modeling the build context through an instruction-level context profiler, a context-aware dependency graph, and a two-step LLM repair workflow.
Evaluated on $D^3$, a benchmark of real-world drift instances, \my achieves a repair rate of 35.22\%, which is 2.78$\times$ that of the rule-based baseline and 1.24$\times$ that of the best LLM-based baseline, while eliminating the prompt-overflow failures that affect all competing methods.
The stateful context simulation and dependency graph representation underlying \my are general abstractions applicable to other sequential IaC languages, pointing toward a broader program of context-aware, AI-assisted IaC maintenance.

\bmsubsection*{Data Availability}

The code and dataset that support the findings of this study are openly available in the Cadre repository on GitHub~\cite{Cadre} at \url{https://github.com/dw763j/Cadre}, reference number [56].






\bmsubsection*{Conflicts of Interest}

The authors declare no conflicts of interest.

\bibliography{ref}




\end{document}